\documentclass[journal,twocolumn,10pt]{IEEEtran}

\usepackage{amssymb}
\usepackage{amsmath}
\usepackage{amsmath,bm}
\usepackage{amsthm}
\usepackage{graphicx}
\usepackage{subfigure}
\usepackage{cite}
\usepackage{enumerate}
\usepackage[ruled]{algorithm2e}
\usepackage{multirow}
\usepackage{color, soul}
\usepackage{url}
\usepackage{enumitem}

\begin{document}

\newtheorem{definition}{\bf Definition}
\newtheorem{theorem}{\bf Theorem}
\newtheorem{lemma}{\bf Lemma}
\newtheorem{assumption}{\bf Assumption}
\newtheorem{remark}{\bf Remark}
\newtheorem{proposition}{\bf Proposition}
\newtheorem{corollary}{\bf Corollary}
\newtheorem{property}{\bf Property}
\newtheorem{problem}{\bf Problem}
\newcommand{\e}[1]{\ensuremath{\times 10^{#1}}}
\newcommand{\ud}{\mathrm{d}}
\newcommand{\unsim}{\mathord{\sim}}
\newcommand{\tabincell}[2]{\begin{tabular}{@{}#1@{}}#2\end{tabular}}

\renewcommand{\arraystretch}{1.3}
\renewcommand{\proofname}{\indent \textbf {Proof}}
\def\QEDclosed{\mbox{\rule[0pt]{1.3ex}{1.3ex}}}
\def\QED{\QEDclosed}
\def\proof{\indent{\textbf{\emph{Proof: }}}}
\def\endproof{\hspace*{\fill}~\QED\par\endtrivlist\unskip}
\renewcommand{\footnoterule}{
    \kern -3pt
    \hrule width 0.488\textwidth
    \kern 2.6pt
}
\SetAlFnt{\footnotesize}
\SetAlCapFnt{\footnotesize}
\SetAlCapNameFnt{\footnotesize}

\title{ImgSensingNet: UAV Vision Guided Aerial-Ground Air Quality Sensing System}

\author{\IEEEauthorblockN{{Yuzhe Yang},~\IEEEmembership{Student Member,~IEEE,}
		{Zhiwen Hu},~\IEEEmembership{Student Member,~IEEE,}\\
		{Kaigui Bian},~\IEEEmembership{Member,~IEEE,}
		and {Lingyang Song},~\IEEEmembership{Fellow,~IEEE}}\\
\thanks{Y. Yang is with the Computer Science and Artificial Intelligence Laboratory, MIT, Cambridge, MA 02139 (email: yuzhe@mit.edu).}
\thanks{Z. Hu, K. Bian and L. Song are with School of Electrical Engineering and Computer Science, Peking University, Beijing, China (email: \{zhiwen.hu, bkg, lingyang.song\}@pku.edu.cn).}
}

\maketitle
\setlength{\abovecaptionskip}{0pt}
\setlength{\belowcaptionskip}{-10pt}

\begin{abstract}
Given the increasingly serious air pollution problem, the monitoring of air quality index~(AQI) in urban areas has drawn considerable attention.
This paper presents \emph{ImgSensingNet}, a vision guided aerial-ground sensing system, for fine-grained air quality monitoring and forecasting using the fusion of haze images taken by the unmanned-aerial-vehicle (UAV) and the AQI data collected by an on-ground three-dimensional~(3D) wireless sensor network~(WSN). Specifically, \emph{ImgSensingNet} first leverages the computer vision technique to tell the AQI scale in different regions from the taken haze images, where haze-relevant features and a deep convolutional neural network~(CNN) are designed for direct learning between haze images and corresponding AQI scale.
Based on the learnt AQI scale, \emph{ImgSensingNet} determines whether to wake up on-ground wireless sensors for small-scale AQI monitoring and inference, which can greatly reduce the energy consumption of the system. An entropy-based model is employed for accurate real-time AQI inference at unmeasured locations and future air quality distribution forecasting.
We implement and evaluate \emph{ImgSensingNet} on two university campuses since Feb. 2018, and has collected 17,630 photos and 2.6 millions of AQI data samples. Experimental results confirm that \emph{ImgSensingNet} can achieve higher inference accuracy while greatly reduce the energy consumption, compared to state-of-the-art AQI monitoring approaches.
\end{abstract}


\IEEEpeerreviewmaketitle

\section{Introduction}
Air pollution has been proved to have significantly negative effects on human health and sustainable development~\cite{who}. Air pollution is caused by gaseous pollutants that are harmful to humans and ecosystem.
To quantify the degree of air pollution, government agencies have defined the air quality index~(AQI). AQI is calculated based on the concentration of a number of air pollutants, such as ${\rm PM}_{2.5}$ and ${\rm PM}_{10}$ particles.
A higher AQI indicates that air pollution is more severe and people are more likely to experience harmful health effects~\cite{station}. Thus, AQI monitoring is a critical issue. The more accurate AQI distribution that can be obtained in a region, the more effective methods we can find to deal with the air pollution.

Existing AQI monitoring approaches can be classified into two categories. The first category includes the \emph{sensor-based monitoring} approaches, wherein government agencies have
set up monitoring stations on dedicated sites in a city~\cite{official station}. However, these fixed stations only provide coarse-grained 2D monitoring, with several kilometers between two monitoring stations. Existing study has shown that AQI distribution has intrinsic variation within meters~\cite{whyfineaqi}. Large scale Internet-of-Things~(IoT) applications have been developed to monitor the fine-grained air quality using densely deployed sensors~\cite{aircloud,mosaic}. Although the static sensors may achieve the high precision of monitoring, they suffer from the high cost as well as lack of mobility. Mobile devices or vehicles, such as phones, cars, balloons are utilized to carry sensors for AQI monitoring~\cite{balloon,arms,aqnet,blueaer}. However, the sensor-based approach may induce high energy consumptions for mobile devices or vehicles to acquire certain amount of data.

The second category of approaches includes the \emph{vision-based monitoring}. Image-based AQI monitoring stations are set up by researchers at dedicated locations~\cite{ibaqms}, and these static stations can only take photos and infer the AQI at limited sites over the whole region. Crowd-sourced photos contributed by mobile phones can depict the AQI distribution~\cite{cell phone} at more locations. However, the performance of the crowd sourcing approach is usually restricted by the low quality photos contributed by many non-savvy users.


\begin{table*}[!htbp]
\caption{Comparison of State-of-the-art Air Quality Monitoring Approaches}
\centering
\begin{tabular}{ c|c|c|c|c|c|c|c|c }
\hline
Systems & Scale & Dimension & \tabincell{c}{Monitoring\\ Methods} & Resolution & Mobility & Costs & \tabincell{c}{Real-Time\\ Capability} & Accuracy \\
\hline\hline
Official stations~\cite{official station} & $\sim{100}$ km & $2$-D & Sensor & Low & Static & High & No & Low \\
\hline
AirCloud~\cite{aircloud} & $\sim{5}$ km & $2$-D & Sensor & Medium & Static & Low & No & Medium \\
\hline
Mosaic~\cite{mosaic} & $\sim{5}$ km & $2$-D & Sensor & Medium & Mobile & Low & No & Medium \\
\hline
Mobile nodes~\cite{mobile node} & $1$ km & $2$-D & Sensor & Medium & Mobile & Medium & Yes & Medium \\
\hline
Balloons~\cite{balloon} & $1$ km & $3$-D & Sensor & High & Mobile & Medium & No & Low \\
\hline
BlueAer~\cite{blueaer} & $\sim{10}$ km & $3$-D & Sensor & Medium & Static+Mobile & High & No & High \\
\hline
ARMS~\cite{IoTJ} & $100$ m & $3$-D & Sensor & High & Mobile & High & Yes & High \\
\hline
AQNet~\cite{aqnet} & $\sim{2}$ km & $3$-D & Sensor & High & Static+Mobile & Low & Yes & High \\
\hline\hline
Cell phones~\cite{cell phone} & $4$ km & $2$-D & Vision & Low & Mobile & Low & Yes & Medium \\
\hline
IBAQMS~\cite{ibaqms} & $\sim{1}$ km & $2$-D & Vision & Low & Static & Medium & No & Medium \\
\hline
\textbf{ImgSensingNet} & \textbf{\boldmath{$\sim{10}$} km} & \textbf{\boldmath{$3$-D}} & \textbf{Sensor+Vision} & \textbf{High} & \textbf{Static+Mobile} & \textbf{Low} & \textbf{Yes} & \textbf{High} \\
\hline
\end{tabular}
\label{table:related work}
\end{table*}

Previous works have separated the two categories of methods in AQI monitoring; however, \emph{sensor-based} and \emph{vision-based} methods can be combined to promote the performance of the mobile sensing system, while reducing the power consumption. For example, the combination of computer vision and inertial sensing has been proved to be successful in the task of localization and navigation by phones~\cite{kshin mobicom,sextant}. In this work, we seek a way of leveraging both photo-taking and data sensing to monitor and infer the AQI value.

In this paper, we present \emph{ImgSensingNet}, a UAV vision guided aerial-ground air quality sensing system, to monitor and forecast AQI distributions in spatial-temporal perspectives. Unlike existing systems, we implement: (1) mobile vision-based sensing over an unmanned-aerial-vehicle~(UAV), which realizes three-dimensional~(3D) AQI monitoring by UAV photo-taking instead of using particle sensors, to infer region-level AQI scale (an interval of possible AQI values) by applying a deep convolutional neural network~(CNN) over the taken hazy photos; (2) ground sensing over a wireless sensor network~(WSN) for small-scale accurate spatial-temporal AQI inference, using an entropy-based inference model; (3) an energy-efficient wake-up mechanism that powers on the sensors in a region when small-scale monitoring is needed in that region, based on the result of vision-based AQI inference, which greatly reduces energy consumption while maintaining high inference accuracy.
We implement and evaluate ImgSensingNet on two university campuses~(i.e., Peking University and Xidian University) since Feb. 2018. We have collected 17,630 photos and 2.6 millions of data samples.
Compared to state-of-the-art methods, evaluation results confirm that ImgSensingNet can save the energy consumptions by 50.5\% while achieving an accuracy of 95.2\% for inference.

The main contributions are summarized as below.
\begin{itemize}
\item[$\bullet$] We implement ImgSensingNet, a UAV vision guided aerial-ground AQI sensing system, and we deploy and evaluate it in the real-world testbed;
\item[$\bullet$] The proposed vision-based sensing method can learn the direct correlation between raw haze images and corresponding AQI scale distribution;
\item[$\bullet$] The proposed entropy-based inference model for ground WSN can achieve a high accuracy in both real-time AQI distribution estimation and future AQI prediction;
\item[$\bullet$] The wake-up mechanism connects the aerial vision technique with the on-ground WSN, which can greatly save the energy consumptions of the on-ground sensor network while ensuring high inference and prediction accuracy
\end{itemize}

The rest of this paper goes as follows. Related works are introduced in Section II. In Section III, we present the system overview of ImgSensingNet. Section IV introduces the UAV vision-based aerial sensing. In Section V, we propose the AQI inference model for ground WSN. Section VI introduces the energy-efficient wake-up mechanism. In Section VII, we detail the system implementation. Experimental results and conclusions are provided in Section VIII and Section IX.

\section{Taxonomy}

\subsection{AQI Monitoring Methods}
In Table~\ref{table:related work}, we show state-of-the-art works on air quality monitoring systems. Existing AQI monitoring methods can be summarized into two categories.

\textbf{Sensor-based:} Stationary stations~\cite{station} are set up on dedicated sites in a city, but only provide a limited number of measurement samples. For example, there are only 28 monitoring stations in Beijing. The distance between two nearby stations is typically several ten-thousand meters, and the AQI is monitored every 2 hours~\cite{official station}. AirCloud~\cite{aircloud} uses densely distributed sensors in a static way, while Mosaic~\cite{mosaic} and~\cite{mobile node,balloon} adopt mobile devices such as buses or balloons to carry low-cost sensors. However, they all fail to consider the heterogeneous 3D AQI distribution. In \cite{blueaer,aqnet,arms,ieee-network}, drones with sensors together with ground sensors are used for AQI profiling. However, they are either restricted in a small scale region or may induce high costs, without designing energy-efficient schemes for integrating aerial sensing with ground sensing.

\textbf{Vision-based:} Instead of various particle sensors, image-based approaches are also used for AQI estimation. In~\cite{ibaqms}, image-based air quality monitoring stations are set up at dedicated sites over a city. Again, these methods can only profile AQI at a limited number of locations. In~\cite{cell phone}, camera-enabled mobile devices are used for generating crowd-sourced photos for AQI monitoring. However, the incentive to stimulate users for volunteer high-quality photo-taking is the pain point for such a crowd-sourced system. Without precise correlations between haze images and AQI values, they cannot generalize well and may introduce low accuracy.

ImgSensingNet overcomes the above shortcomings by using vision guided aerial sensing to extend sensing scope, while also combining it with ground WSN for accurate AQI distribution inference. An energy-efficient wake-up mechanism is designed to switch on or off the on-ground WSN by examining the aerial sensing results, which greatly lowers the system’s energy consumption.

\subsection{AQI Inference at Unmeasured Locations}
In real-world sensing applications, it is not feasible to acquire AQI data samples at all locations within a region. Hence, AQI modeling and inference are used to estimate AQI distributions at unmeasured locations. Again, the inference models can be summarized into two categories. 

\textbf{Inference by sensor data:} Zheng et al.~\cite{uair} propose to infer air quality based on data from official air quality stations and other features such as the meteorological data. In~\cite{aircloud,mosaic}, crowd-sourcing and a Gaussian process model are used for 2D AQI inference. \cite{blueaer} extends the inference to 3D space by using a random walk model. A fine-grained AQI distribution model is proposed in~\cite{IoTJ} for real-time AQI estimation over a 3D space. Long-short term memory~(LSTM) networks are used in~\cite{lstm} to utilize historical data for more accurate inference. To do the temporal inference, neural networks~(NN) are used~\cite{aqnet} to analyze spatial-temporal correlations and to forecast future distribution.

\textbf{Inference by image data:} Image-based inference has been used to estimate AQI from haze images by designing appropriate inference models. Classical image processing methods as well as machine learning techniques are used in~\cite{ibaqms,cell phone} to model the correlation between haze images and the degree of air pollution.

In this work, we investigate two novel inference models: (1) image-based AQI scale inference in different monitoring regions by computer vision, and (2) the fine-grained spatial-temporal AQI value inference at locations inside each region by ground WSN.

\begin{figure}[!htbp]
\centering
\includegraphics[width=0.5\textwidth]{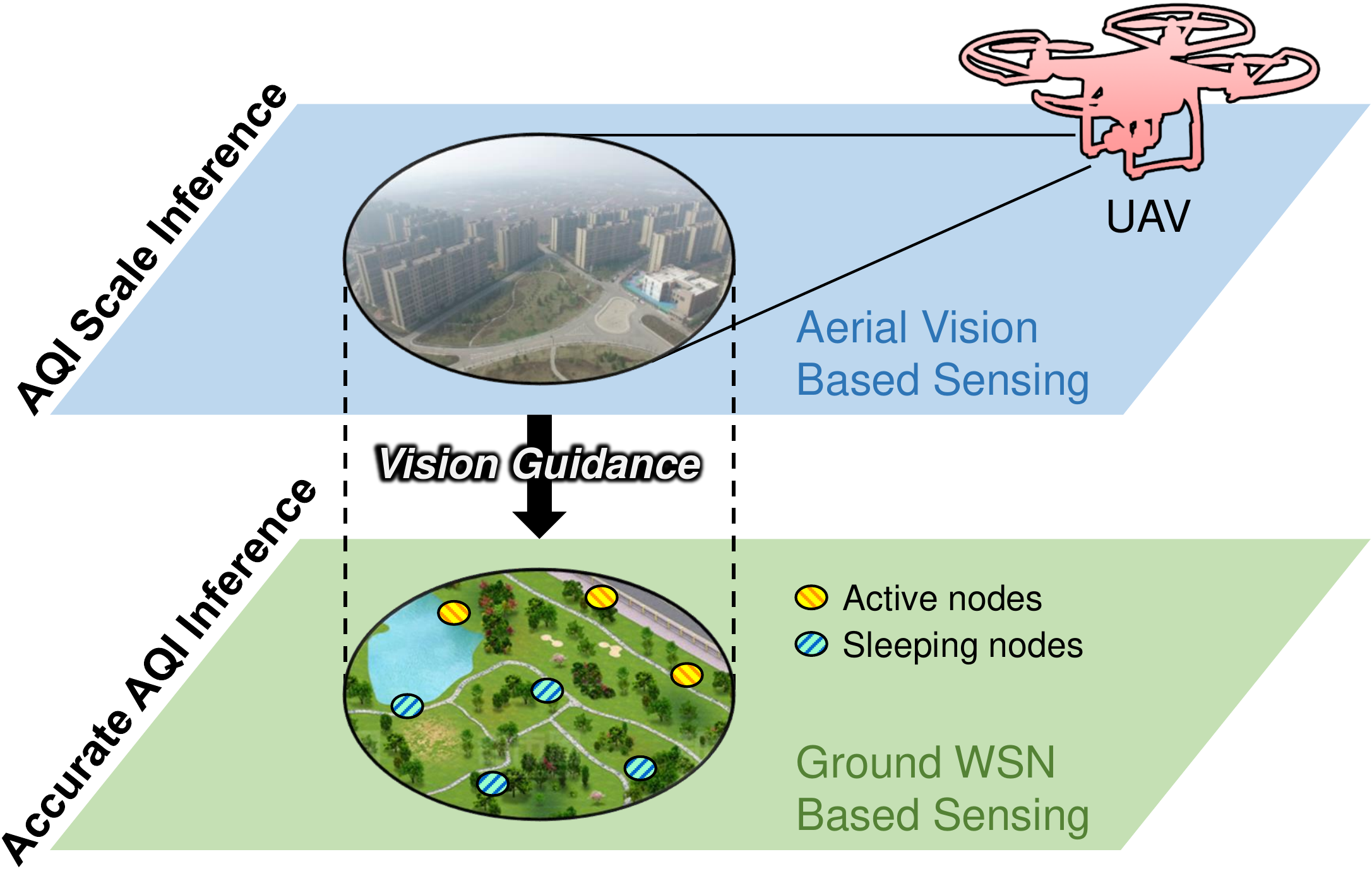}
\caption{The overall framework of \emph{ImgSensingNet}.}
\label{fig:overall framework}
\end{figure}

\section{System Overview}
The ImgSensingNet system includes on-ground programmable monitoring devices and a UAV.
The aerial UAV sensing and the ground WSN sensing form a hybrid sensing network, as illustrated in Fig.~\ref{fig:overall framework}.

The central idea of ImgSensingNet is to trigger aerial sensing and ground sensing sequentially during one measurement, which can provide coarse-to-fine grained AQI value inference. This operation can not only achieve high accuracy, but also scale down the monitoring overhead, which can guarantee a long battery duration without external power supply.

\subsection{Aerial Sensing}
Fig.~\ref{fig:overall framework} shows the overall framework of ImgSensingNet. The aerial sensing utilizes the UAV camera to capture a series of haze images in different monitoring regions. The raw image data is streamed back to the central server, where a well-trained deep learning model performs real-time image data analysis and output the inferred AQI scale for each region.

\subsection{Ground Sensing}
Ground WSN adopts a spatial-temporal inference model for AQI estimation at unmeasured locations and future air quality prediction. Every time when aerial sensing is finished, each ground device follows a designed wake-up mechanism to decide whether to wake up for data sensing based on both the inference result at last time and the aerial sensing result. In this way, the real-time fine-grained AQI distribution is obtained and the future distribution can also be forecasted.

\begin{figure*}[!t]
\centering
\subfigure[Origin image]{
    \label{Origin image}
    \includegraphics[width=0.135\textwidth]{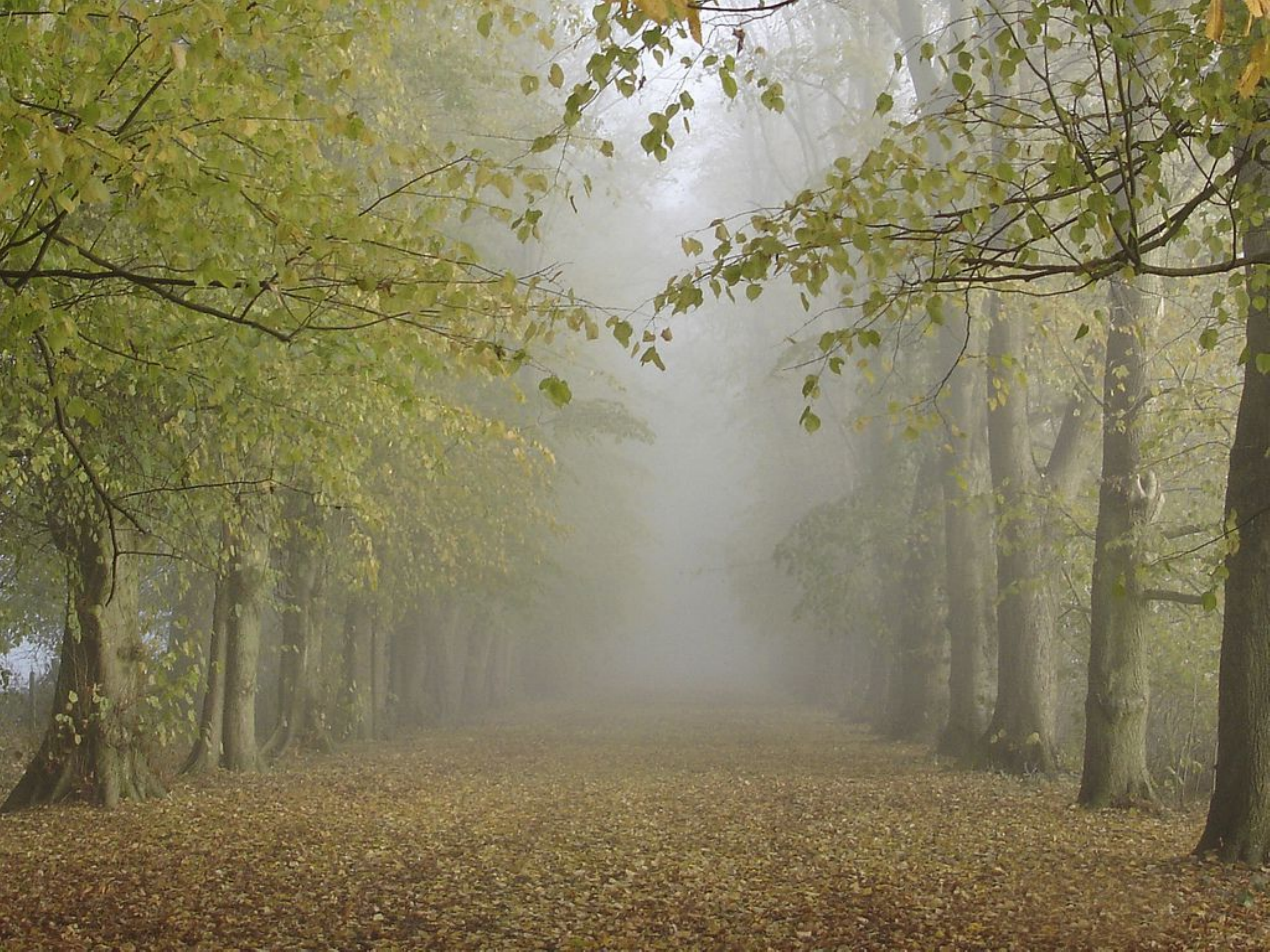}
}
\hspace{-2.2ex}
\subfigure[$F_1$]{
    \label{Refined dark channel}
    \includegraphics[width=0.135\textwidth]{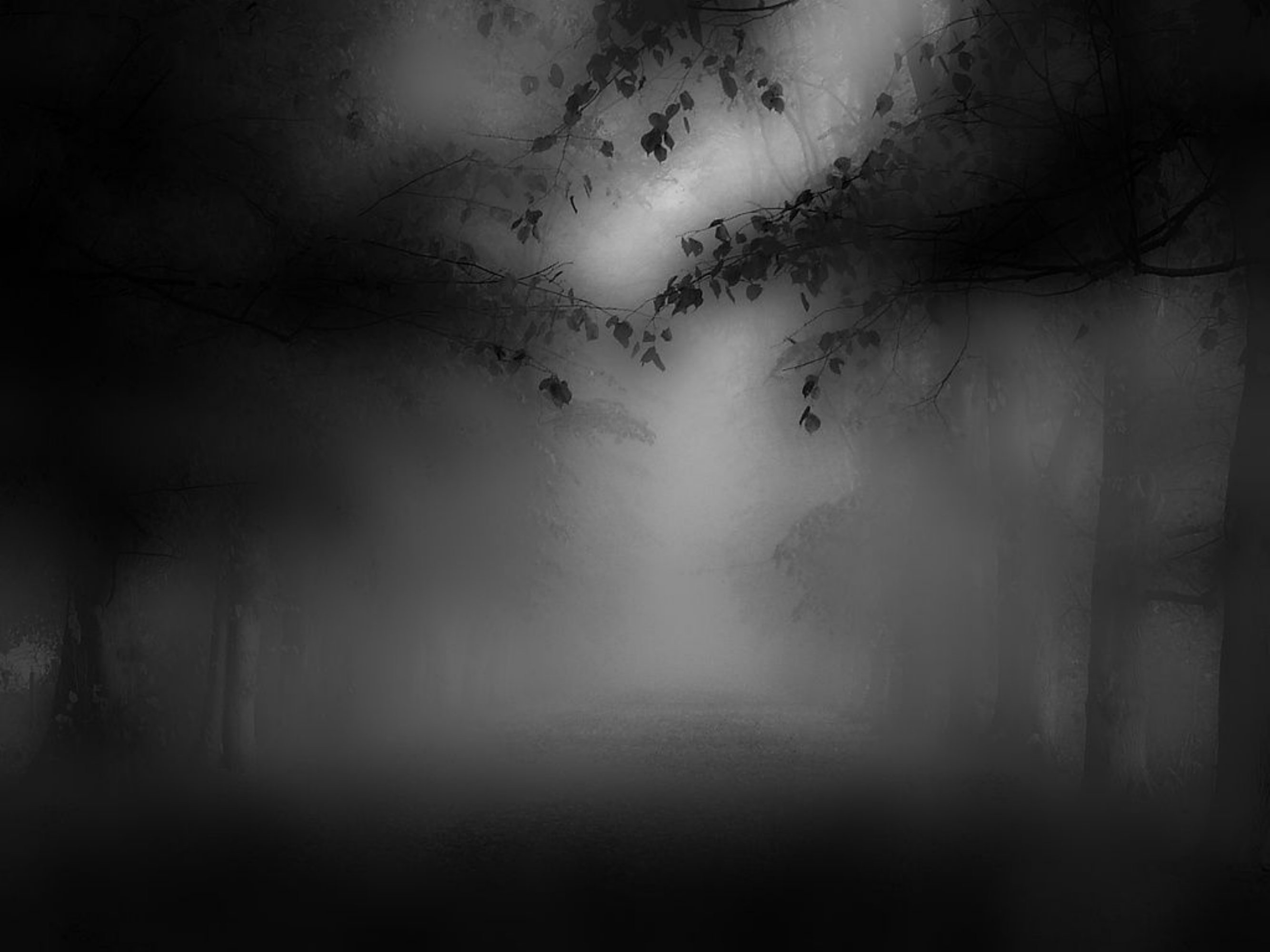}
}
\hspace{-2.2ex}
\subfigure[$F_2$]{
    \label{Max local contrast}
    \includegraphics[width=0.135\textwidth]{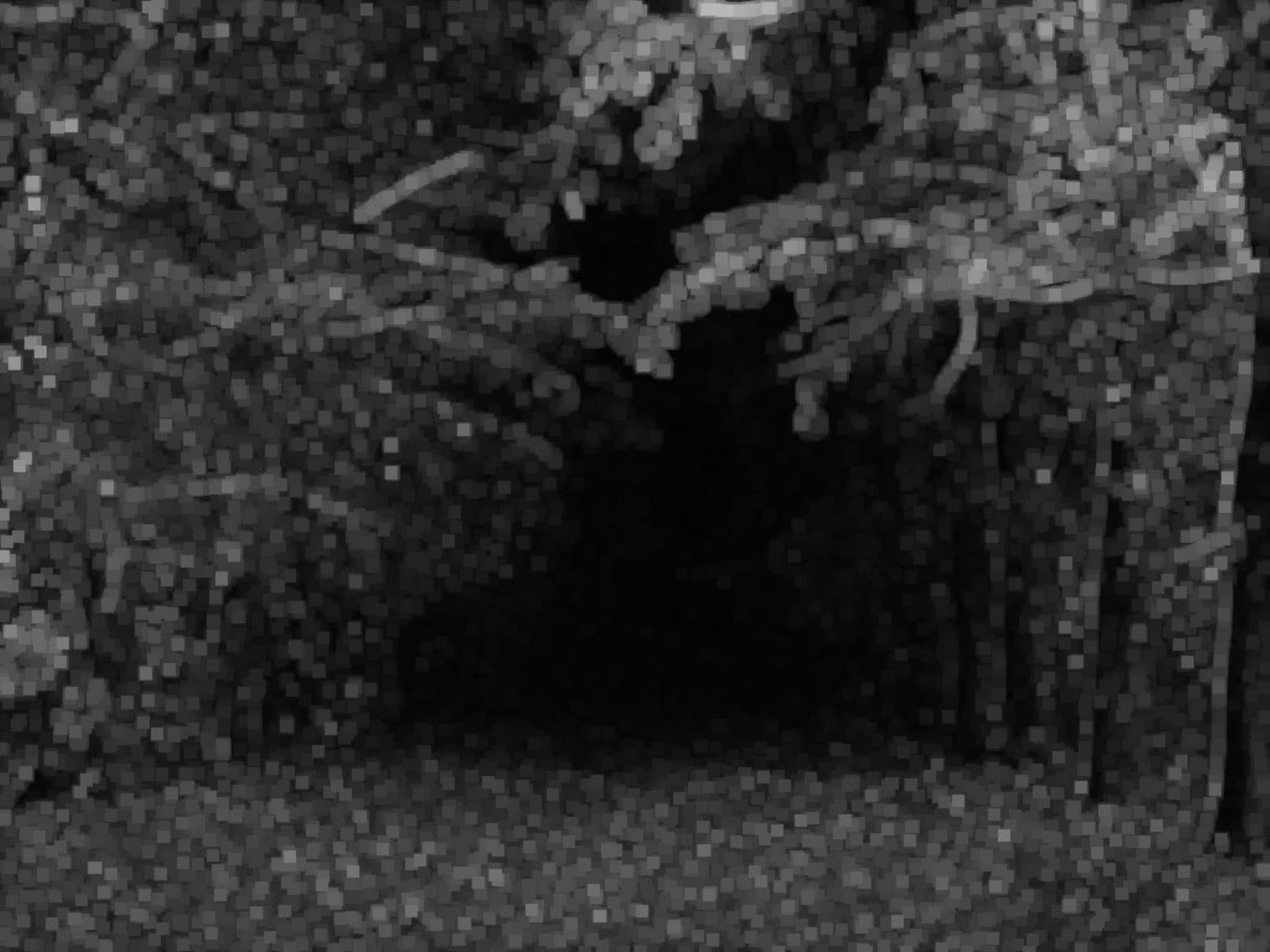}
}
\hspace{-2.2ex}
\subfigure[$F_3$]{
    \label{Max local saturation}
    \includegraphics[width=0.135\textwidth]{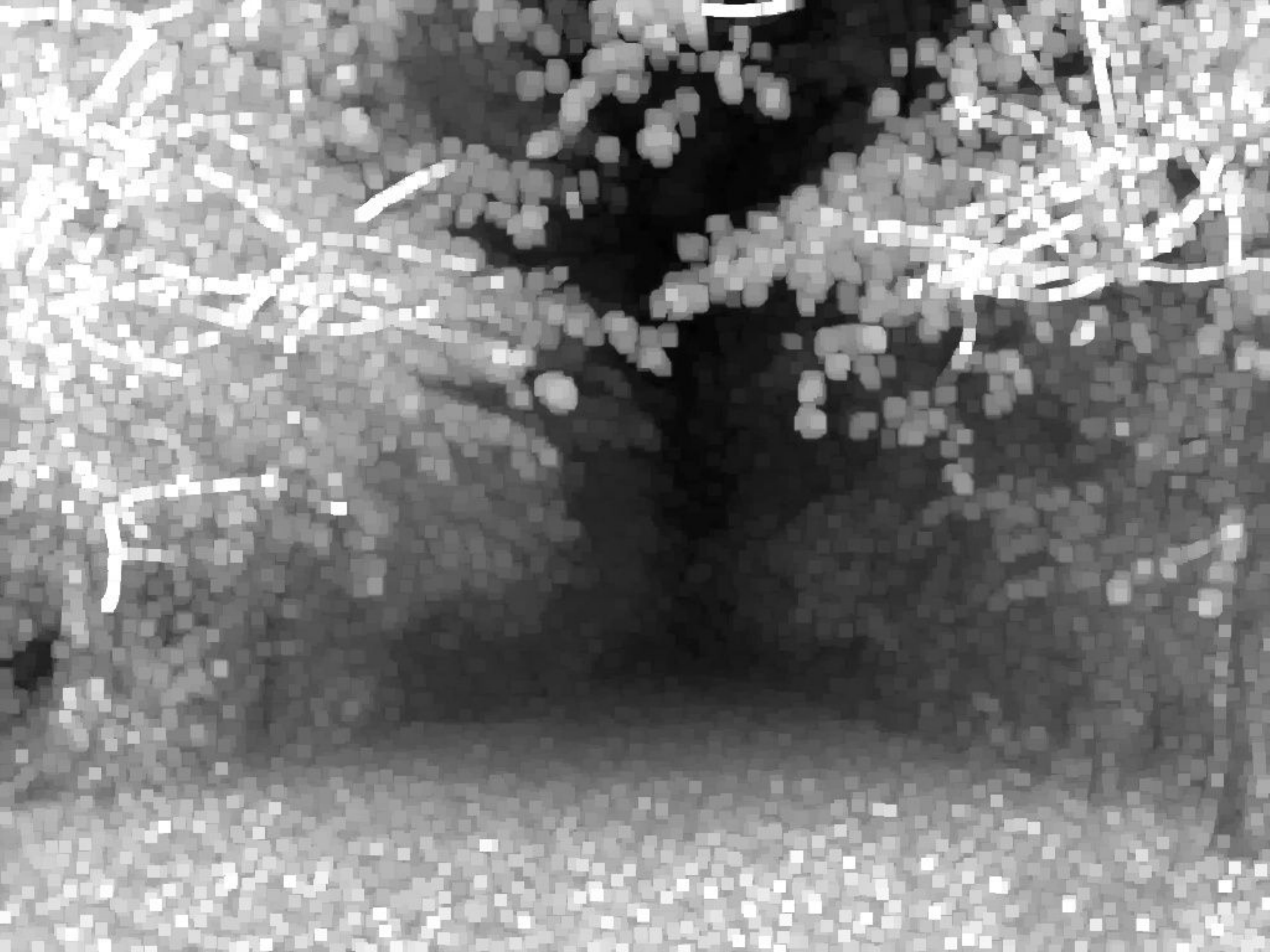}
}
\hspace{-2.2ex}
\subfigure[$F_4$]{
    \label{Min local color attenuation}
    \includegraphics[width=0.135\textwidth]{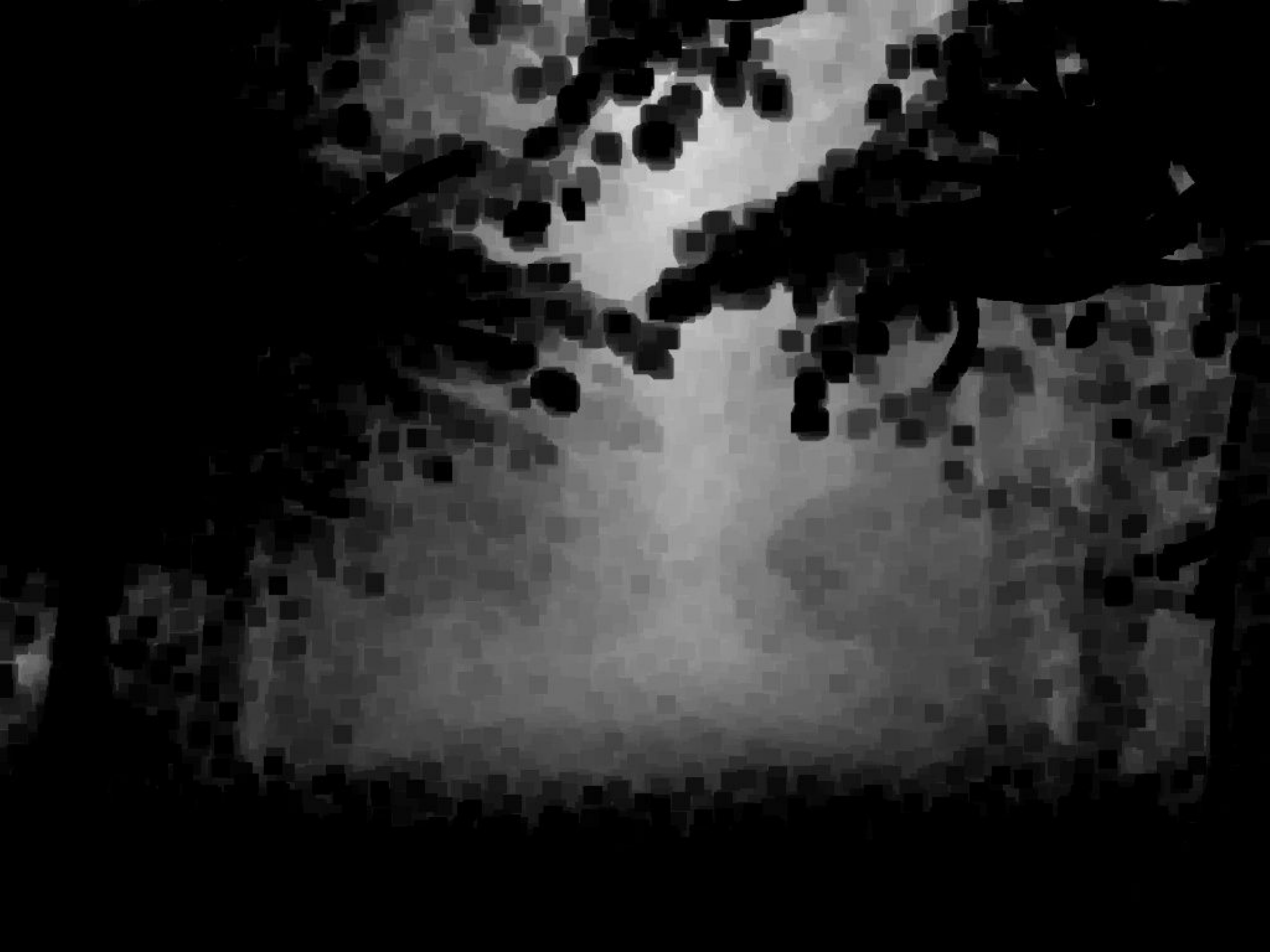}
}
\hspace{-2.2ex}
\subfigure[$F_5$]{
    \label{Hue disparity}
    \includegraphics[width=0.135\textwidth]{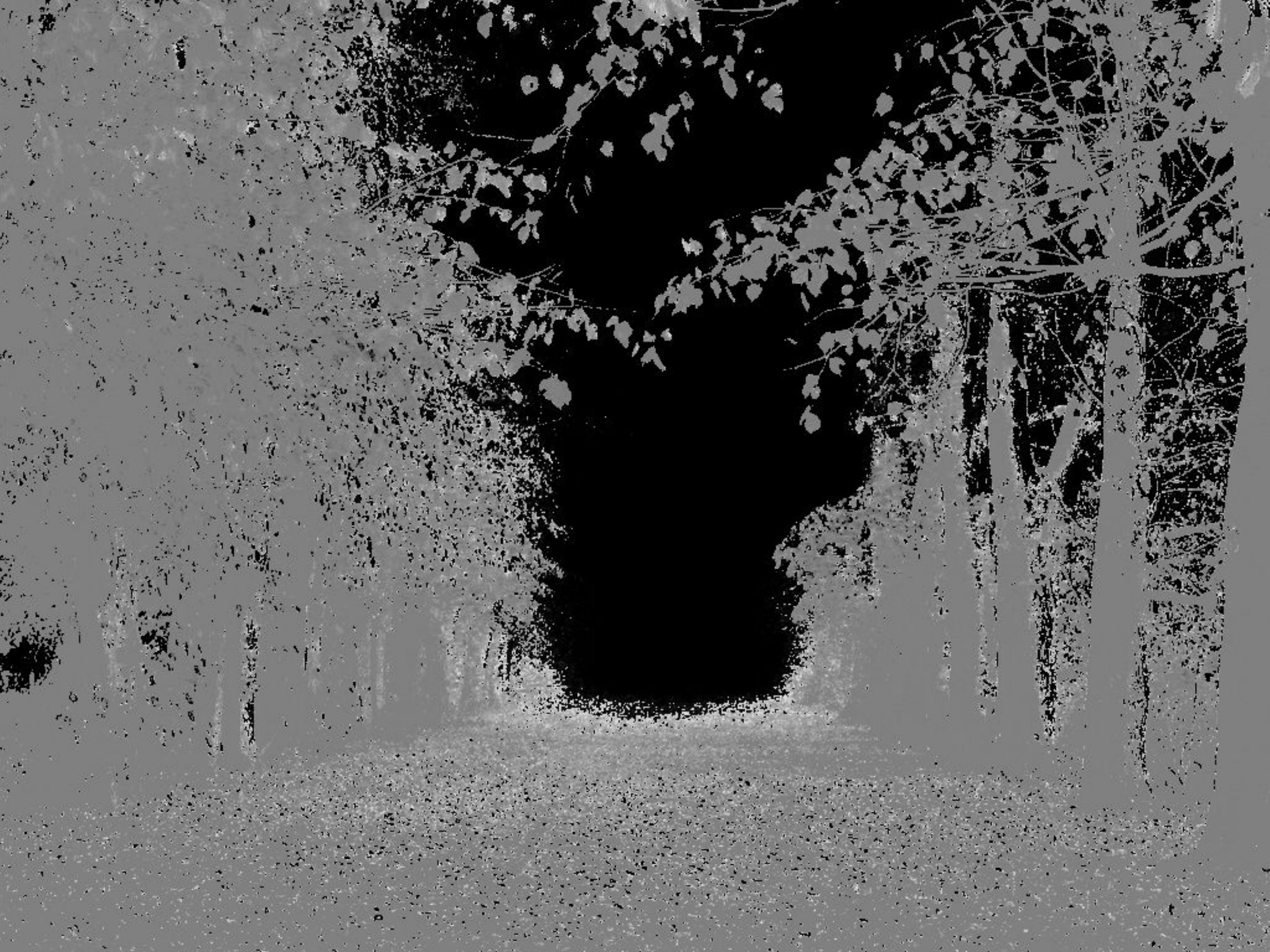}
}
\hspace{-2.2ex}
\subfigure[$F_6$]{
    \label{Chroma}
    \includegraphics[width=0.135\textwidth]{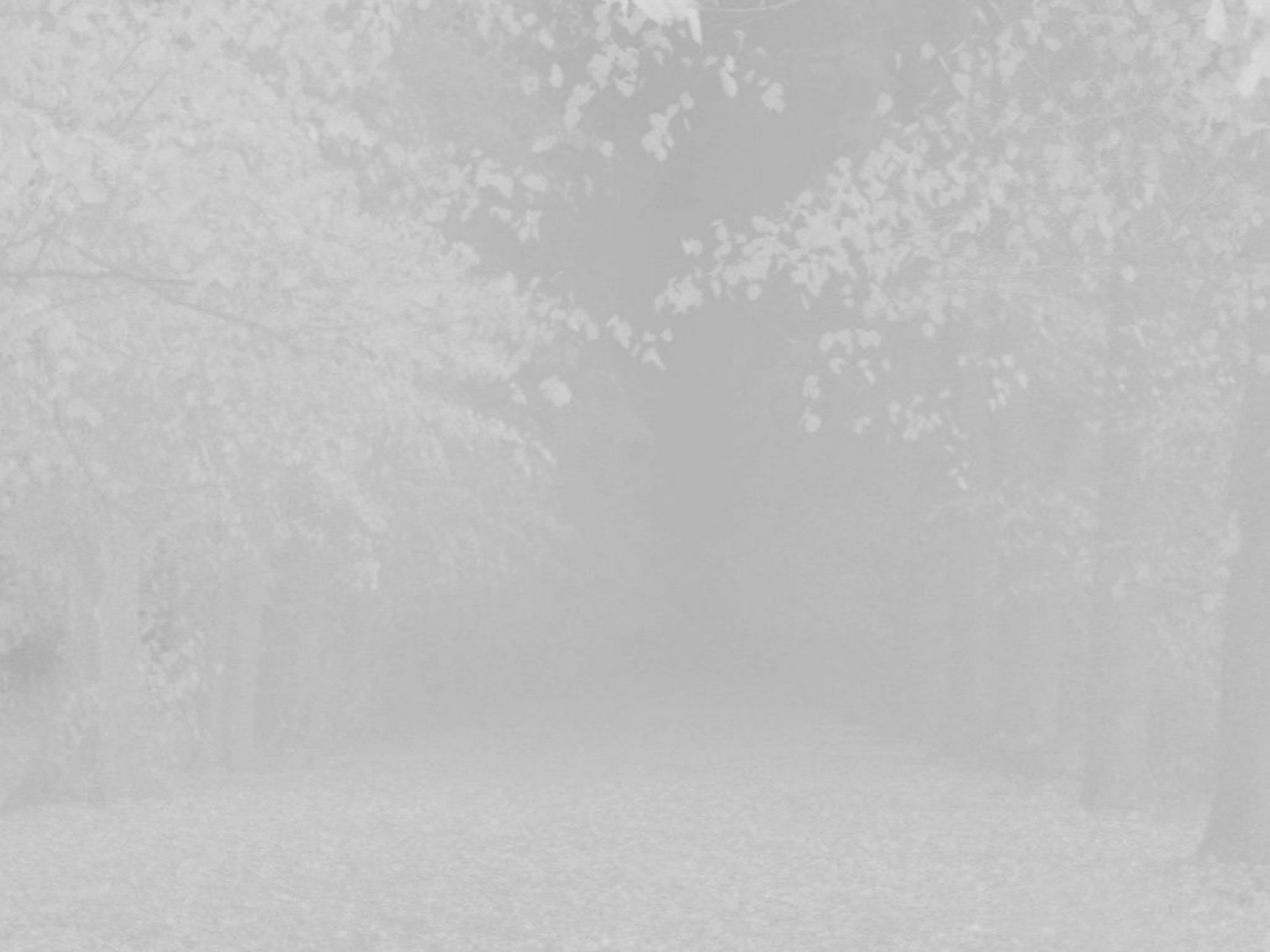}
}
\caption{An example of the extracted feature maps for raw haze image: (a) origin input image; (b) refined dark channel; (c) max local contrast; (d) max local saturation; (e) min local color attenuation; (f) hue disparity; (g) chroma.}
\label{fig:haze features}
\vspace{-0.3cm}
\end{figure*}

\section{Aerial Sensing: Learning AQI Scale from Images Captured by UAV}
ImgSensingNet performs vision-based sensing using UAV, because: (1) the UAV has intrinsic advantages in flexible 3D space sensing over different heights and angles, which avoids possible obstacles, and also guarantees certain scene depths; (2) with built-in camera, the UAV does not need to carry extra sensors, which enables longer monitoring time; and (3) instead of hovering at different locations to collect data by sensors, the UAV can keep flying and video recording by cameras through monitoring regions, which greatly extends the sensing scope.

Recent works have well studied how to remove haze from images in the computer vision field~\cite{attenuation prior,hue,max local contrast,dark channel}. However, there has not been works on quantifying the haze in the image to real AQI value. To do the direct learning from raw haze images to quantified AQI values, two main problems should be answered: \emph{(1) how to extract the haze components from origin images to eliminate the influence of image content}, and \emph{(2) how to quantify the AQI based on the haze components}.

This section details the method to solve these problems. Specifically, we investigate content-nonspecific haze-relevant features for raw haze images. With the haze features extracted, a novel 3D CNN model is designed to better process feature maps and output the inferred AQI scale for each single image.

\subsection{Overview of Haze Image Processing}
In image processing, a haze image can be mathematically described using the haze image formation model~\cite{dark channel} as
\begin{equation}
\bm{\mathcal{I}}\left(\bm{x}\right) = \bm{\mathcal{J}}\left(\bm{x}\right) t\left(\bm{x}\right) + \bm{L_{\infty}}\left( 1 - t\left(\bm{x}\right) \right),
\label{equation:haze model}
\end{equation}
where $\bm{\mathcal{I}}$ is the observed hazy image, $\bm{\mathcal{J}}$ is the haze-free image, $t$ denotes the medium transmission, $\bm{L_{\infty}}$ is the global atmospheric light, and $\bm{x}$ represents pixel coordinates. The haze-removal methods have spent large effort estimating $\bm{\mathcal{J}}$ and $t$ for haze-free image recovery~\cite{attenuation prior,hue,max local contrast,dark channel}.

Instead, in this work we propose a new objective to estimate the degree of haze in a single image.

\subsection{Haze-relevant Features Extraction}
The first step is to extract a list of haze-relevant statistical features. Since we want to investigate general approach for all image inputs regardless of their contents, the features that correlate well with haze density in images but do not correlate well with image contents should be selected.

In the following, we investigate six content-nonspecific haze-relevant features, and an example is illustrated in Fig.~\ref{fig:haze features}.

\subsubsection{Refined Dark Channel}
Dark channel~\cite{dark channel} is an informative feature for haze detection, defined as the minimum of all pixel colors in a local patch:
\begin{equation}
D \left( \bm{x};\bm{\mathcal{I}} \right) = \min_{\bm{y}\in\Omega\left(\bm{x}\right)} \left( \min_{c\in \left\{ r, g, b\right\} } \frac{\mathcal{I}^{c} \left( \bm{y}\right) }{L^c_{\infty}} \right),
\end{equation}
where $\Omega\left(\bm{x}\right)$ is a local patch centered at $\bm{x}$, $\mathcal{I}^{c}$ is one color channel of $\bm{\mathcal{I}}$. It is found that most local patches in outdoor haze-free images contain some pixels whose intensity is very low in at least one color channel~\cite{dark channel}. Therefore, the dark channel is a rough approximation of the thickness of the haze.

To obtain a better estimation of haze density, we propose the \emph{refined dark channel} by applying the guided filter~\cite{guided filter} $\mathcal{G}$ on the estimated medium transmission $\tilde{t}$, to capture the sharp edge discontinuous and outline the haze profile. Note that by applying the min operation on (\ref{equation:haze model}), the dark channel of $\bm{\mathcal{J}}$ tends to be zero, and we have $\tilde{t}(\bm{x}) = 1 - D \left( \bm{x};\bm{\mathcal{I}} \right)$. Hence, the refined dark channel can be expressed as
\begin{equation}
D^{R} \left( \bm{x};\bm{\mathcal{I}} \right) = 1 - \mathcal{G}\left(1 - \min_{\bm{y}\in\Omega\left(\bm{x}\right)} \left( \min_{c} \frac{\mathcal{I}^{c} \left( \bm{y}\right) }{L^c_{\infty}} \right) \right).
\end{equation}
Fig.~\ref{fig:haze features}(b) shows the refined dark channel feature. As we can see, the feature has a high correlation to the amount of haze in the image.

\subsubsection{Max Local Contrast}
Since haze can scatter the light reaching cameras, the contrast of the haze image can be highly reduced. Therefore, the contrast is one of the most perceived features to detect haze in the scene. The local contrast is defined as the variance of pixel intensities in a local $r\times r$ region compared with the center pixel. Inspired by \cite{max local contrast}, we further use the local maximum of local contrast values in a local patch $\Omega\left(\bm{x}\right)$ to form the \emph{max local contrast} feature as
\begin{equation}
C^{T} \left( \bm{x};\bm{\mathcal{I}} \right) = \max_{\bm{y}\in\Omega\left(\bm{x}\right)} \sqrt{\frac{\sum_{\bm{z}\in\Omega_r\left(\bm{y}\right)} \left\| \bm{\mathcal{I}}\left(\bm{z}\right) - \bm{\mathcal{I}}\left(\bm{y}\right) \right\|^2}{\pi \left| \Omega_r\left(\bm{y}\right) \right|}},
\end{equation}
where $\left| \Omega_r\left(\bm{y}\right) \right|$ denotes the size of the local region $\Omega_r\left(\bm{y}\right)$, and $\pi$ is a constant that equals to the number of channels. Fig.~\ref{fig:haze features}(c) shows the contrast feature, in which the correlation between haze and the contrast feature are visually obvious.

\subsubsection{Max Local Saturation}
It is observed that the image saturation varies sharply with the change of haze in the scene~\cite{ibaqms}. Therefore, similar to image contrast, we define the \emph{max local saturation} feature that represents the maximum saturation value of pixels within a local patch, written as
\begin{equation}
S \left( \bm{x};\bm{\mathcal{I}} \right) = \max_{\bm{y}\in\Omega\left(\bm{x}\right)} \left( 1 - \frac{\min_c \mathcal{I}^c \left(\bm{y}\right)}{\max_c \mathcal{I}^c \left(\bm{y}\right)} \right).
\end{equation}
The max local saturation feature for the ``forest'' image is shown in Fig.~\ref{fig:haze features}(d), which is also correlated with the haze.

\begin{figure*}[!htbp]
\centering
\includegraphics[width=0.95\textwidth]{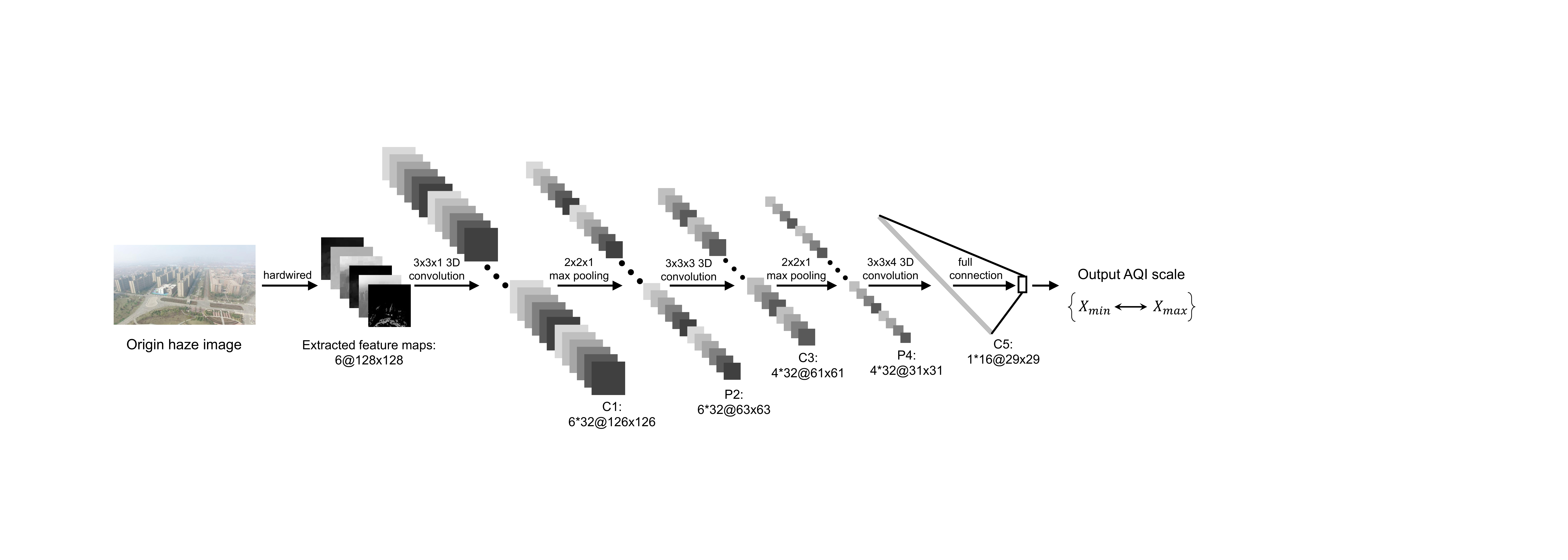}
\caption{The architecture of the proposed 3D CNN model.}
\label{fig:3d cnn model}
\vspace{-0.3cm}
\end{figure*}

\subsubsection{Min Local Color Attenuation}
In~\cite{attenuation prior}, the scene depth is found to be positively correlated with the difference between the image brightness and the image saturation by numerous experiments on haze images. This statistics is regarded as the color attenuation prior, expressed as
\begin{equation}
d \left( \bm{x};\bm{\mathcal{I}} \right) = \theta_0 + \theta_1\cdot \bm{\mathcal{I}}^v\left( \bm{x};\bm{\mathcal{I}} \right) + \theta_2\cdot \bm{\mathcal{I}}^s\left( \bm{x};\bm{\mathcal{I}} \right) + \epsilon \left( \bm{x};\bm{\mathcal{I}} \right),
\end{equation}
where $\bm{\mathcal{I}}^v$ and $\bm{\mathcal{I}}^s$ denote the brightness and the saturation, respectively. Let $\epsilon(\bm{x})\sim \mathcal{N}(0,\sigma^2)$, and $\theta_0$, $\theta_1$, $\theta_2$, $\sigma$ can be estimated through maximum likelihood. To process the raw depth map for better representation of the haze influence, we define the \emph{min local color attenuation} feature by considering the minimum pixel-wise depth within a local patch $\Omega\left(\bm{x}\right)$:
\begin{equation}
A \left( \bm{x};\bm{\mathcal{I}} \right) = \min_{\bm{y}\in\Omega\left(\bm{x}\right)} d \left( \bm{y};\bm{\mathcal{I}} \right).
\end{equation}
Fig.~\ref{fig:haze features}(e) shows the min local color attenuation feature, where an obvious correlation with haze density can be observed.

\subsubsection{Hue Disparity}
In~\cite{hue}, the hue disparity between the original image and its semi-inverse image is utilized to remove haze. The semi-inverse image is defined as the max value between original image and its inverse, expressed as
\begin{equation}
\mathcal{I}^c_{si} \left( \bm{x};\bm{\mathcal{I}} \right) = \max_{\bm{x}\in \bm{\mathcal{I}}} \big\{\ \mathcal{I}^c \left( \bm{x} \right), 1 - \mathcal{I}^c \left( \bm{x} \right) \big\}, \quad c \in \left\{ r, g, b \right\}.
\end{equation}
The \emph{hue disparity} is also reduced by haze, thus can serve as another haze-relevant feature, written as
\begin{equation}
H \left( \bm{x};\bm{\mathcal{I}} \right) = \left| \bm{\mathcal{I}}^h_{si} \left( \bm{x};\bm{\mathcal{I}} \right) - \bm{\mathcal{I}}^h \left( \bm{x};\bm{\mathcal{I}} \right) \right|,
\end{equation}
where $\bm{\mathcal{I}}^h$ denotes the hue channel of the image. Fig.~\ref{fig:haze features}(f) shows the hue disparity feature for haze image.

\subsubsection{Chroma}
In the CIELab color space, the chroma is one of the most representative image feature to describe the color degradation by the haze in the atmosphere. Let $\left[ L\left( \bm{x};\bm{\mathcal{I}} \right)\ a\left( \bm{x};\bm{\mathcal{I}} \right)\ b\left( \bm{x};\bm{\mathcal{I}} \right) \right]^{\rm T}$ denotes the haze image $\bm{\mathcal{I}}$ in the CIELab space, the $chroma$ feature is defined as
\begin{equation}
C^{H} \left( \bm{x};\bm{\mathcal{I}} \right) = \sqrt{a^2\left( \bm{x};\bm{\mathcal{I}} \right) + b^2\left( \bm{x};\bm{\mathcal{I}} \right)}.
\end{equation}
As shown in Fig.~\ref{fig:haze features}(g), chroma is an excellent haze-relevant feature since it strongly correlates with the haze density but is not affected by the image contents.

\subsection{3D CNN-based Learning for AQI Scale Inference}
With the above haze-relevant features extracted, we design a 3D CNN model to perform direct learning for precise AQI scale estimation of input haze images. CNN is a type of deep learning model in which trainable filters and local neighborhood pooling operations are applied alternatively on the raw input images, resulting in a hierarchy of increasingly complex features. CNN has been widely used for image processing and vision applications, and has been proved to achieve superior performance compared to classical methods.

In this work, to better fit the extracted features for high accuracy, we introduce a 3D CNN model by adding a ``prior feature map'' dimension. The advantage behind 3D convolution is the utilization of haze prior information, which is encoded in the six feature maps.

\textbf{Preprocessing:} For each input haze image, we first resize it spatially to $128\times 128$ pixels. The resized image is then performed with feature maps extraction and rescaled into $\left[0,\ 1\right]$ in grayscale. We normalize each dimension except the prior feature map dimension of all training haze images to be of zero mean, which can help our model converge faster.

\textbf{Model Architecture:} Fig.~\ref{fig:3d cnn model} presents the architecture of the 3D CNN model. The first layer is called the ``hardwired'' layer that extracts feature maps from original haze image, consisting of six feature frames stacked together to be a $128\times 128\times 6$ sized tensor. The rationale for using this hardwired layer is to encode our prior knowledge on different haze-relevant features. This scheme regularizes the model training constrained in the prior haze feature space, which leads to better performance compared to random initialization. 

We then apply 3D convolutions with a kernel size of $3\times3\times1$ and 32 kernels, to extract complex features in different feature map domains separately. In the subsequent pooling layer, $2\times2\times1$ max pooling is applied. The next convolution layer uses $3\times3\times3$ kernel size, followed by another $2\times2\times1$ max pooling. 3D convolution with $3\times3\times4$ kernel size is then applied and it contains 13,456 trainable parameters. Finally, the vector is densely connected to the output layer that consists pre-divided AQI scale classes. This architecture has been verified to give the best performance compared to other 3D CNN architectures.

\textbf{Training and AQI Scale Inference:} As the output is AQI scale~(i.e., $\left[ X_{min},\ X_{max} \right]$), the inference is modeled as a classification problem, where the AQI scale classes are pre-divided based on the number of different AQI values in training data. Given new image input, the model finds images in training set with most similar haze degrees, and uses the corresponding AQI ground truth values to generate an AQI scale. With more data of different AQI values collected, the number of class will increase, resulting in more fine-grained scale labels.

\section{Ground Sensing: AQI Inference by Ground Sensor Monitoring}
Given the 3D target monitoring space, we utilize ground WSN for accurate AQI inference that enables both the real-time inference spatially, and future distribution forecasting temporally. This section illustrates how to do accurate inference based on (1) sparse historical ground WSN data, and (2) the prior AQI scale knowledge by aerial sensing.

The target 3D space is first divided into disjointed cubes, which form the basic unit in our inference. Each cube contains its own geographical coordinates in 3D space, and each cube is associated with an AQI value. Note that AQI values in a limited number of cubes are observed/sensed from the WSN, while the AQI values in other unobserved cubes need to be estimated using the proposed model.
Here we define a set of cubes $\{C_1,C_2,\dots,C_s\}$ over a series of time stamps $\{T_1,T_2,\dots,T_d\}$ with equal intervals~(e.g., one hour). Most cubes do not have observed/sensed data~(e.g., $\ge99\%$ in both Peking University and Xidian University), whose AQI values can be estimated using a probability function, $p_u$. The objective is to infer $p_u$ of any unobserved location $C$ at any given time stamp $T_i$~(including both the current and future time stamps).

\textbf{Why a semi-supervised learning model:} Since the data observed using the sensor network can be extremely sparse, prevailing deep learning methods for time series processing~(e.g., RNN and LSTM) are not feasible in our task. Hence, a semi-supervised learning method is designed to achieve the goal. We first establish a multi-layer spatial-temporal graph to model the correlation between cubes. The weights of edges are represented by the correlations of features between cubes, based on the fact that cubes whose features are similar tend to share similar AQI values. The model iteratively learns and adjusts the edge weights to achieve the inference.

\subsection{Feature Selection}
Based on the study for key features in fine-grained scenarios~\cite{arms,aqnet,ieee-network,IoTJ}, we select nine highly correlated features as: \emph{3D coordinates}, \emph{current time stamp}, \emph{weather condition}, \emph{wind speed}, \emph{wind direction}, \emph{humidity} and \emph{temperature}. These features can be obtained either by our monitoring devices or crawling data from online websites.

\subsection{Multi-Layer Spatial-Temporal Inference Model}
The AQI values at different locations are correlated with each other in a spatial-temporal manner. For example, the AQI value at one location is highly similar to that at its neighboring location; the AQI value at a location depend on its values in past few hours.

\begin{figure}[!htbp]
\centering
\includegraphics[width=0.48\textwidth]{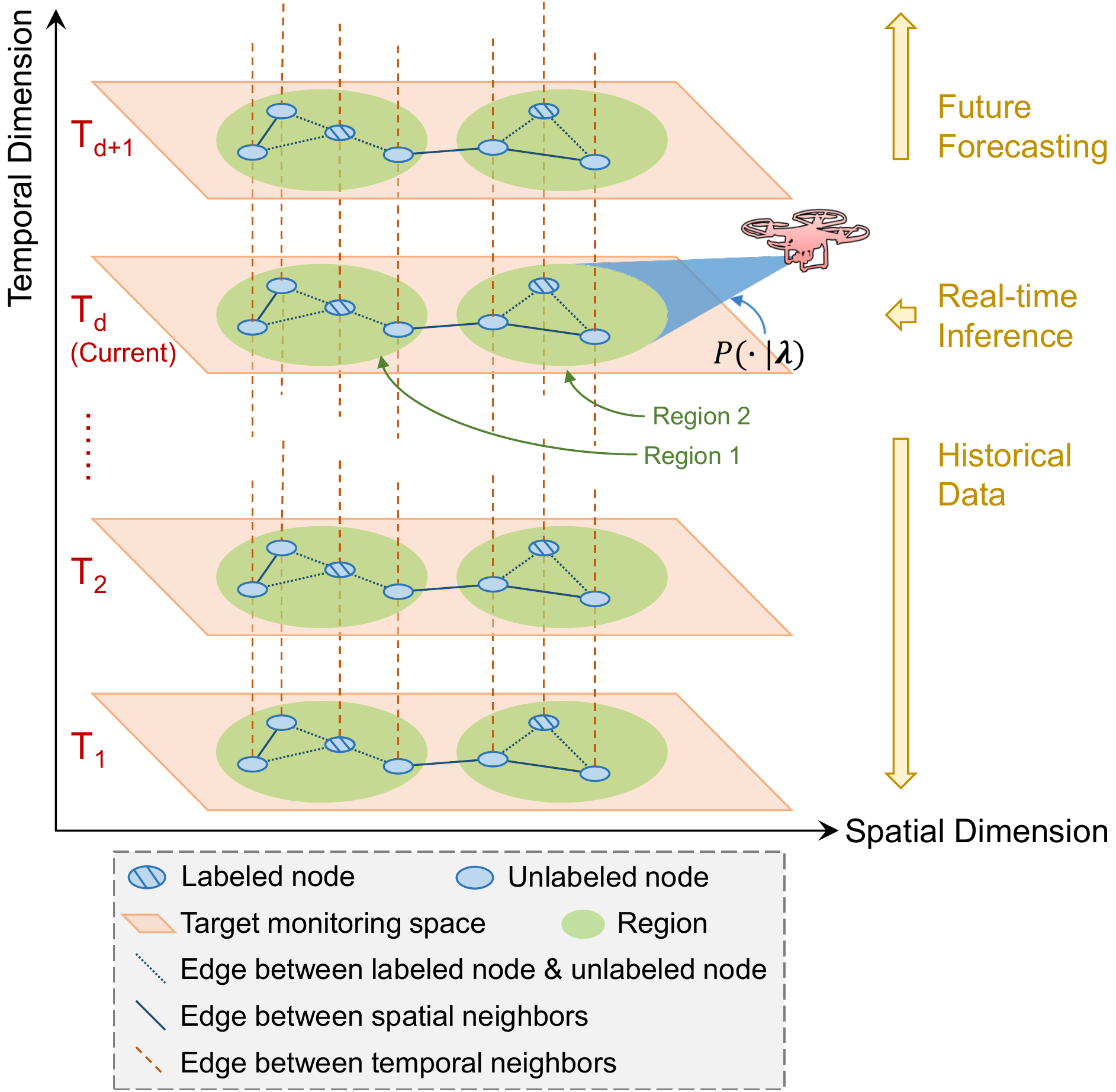}
\caption{An illustration of the proposed multi-layer spatial-temporal correlation graph model.}
\label{fig:multi-layer graph model}
\end{figure}

Based on this observation, we propose a multi-layer graph model to characterize the correlations between cubes. Each cube is represented by a node in the graph, as shown in Fig.~\ref{fig:multi-layer graph model}. These nodes are connected in both spatial and temporal dimensions to form a multi-layer weighted graph $\mathcal{G}=(\mathcal{V},\mathcal{E})$. Each layer represents one spatial graph at a specific time stamp $T_k$. We name the nodes with observed data from the sensors as $labeled$ nodes, while nodes without observed data as $unlabeled$ nodes. Each labeled node $l$ has the ground truth AQI value, while the AQI value of each unlabeled node $u$ is estimated through a probability distribution $p_u$.

We construct the edges $\mathcal{E}$ in the graph by following steps: \textbf{(1) Connecting to labeled nodes,} where each unlabeled node is connected with all labeled nodes at the same time stamp $T_k$; \textbf{(2) Connecting to spatial neighbors,} where each unlabeled node is also connected with neighboring nodes within a given spatial radius $r$; and \textbf{(3) Connecting to temporal neighbors,} where each unlabeled node is connected to nodes in the same location but at neighboring time stamps. Fig.~\ref{fig:multi-layer graph model} shows an example of edge construction.

For every edge $(v_1,v_2) \in \mathcal{E}$, it has a corresponding weight. The weight of edge denotes how much the features between $v_1$ and $v_2$ are correlated. The correlation is defined by:
\begin{definition}
{\rm \textbf{Correlation Function.}} Given a set of features $\textbf{e} = \{e^{(1)},e^{(2)},\dots,e^{(M)}\}$, the correlation function of each feature between node $v_1$ and $v_2$ is defined as a linear function
\begin{align}
Q_{e^{(m)}}(v_1,v_2) =\ & \alpha_m + \beta_m \left\| e^{(m)}(v_1)-e^{(m)}(v_2) \right\|_{1}, \nonumber \\[3pt]
m =\ & 1, 2, \dots, M.
\label{correlation func}
\end{align}
\end{definition}

\noindent In (\ref{correlation func}), $\alpha_m$ and $\beta_m$ are parameters that can be estimated using the maximum likelihood estimation. Based on the correlation modeling between feature difference and AQI similarity, we define the weight matrix $\mathcal{W}=\{w_{i,j}\}$, where the weight on edge $\{(v_1,v_2) \in \mathcal{E}\}$ is expressed as
\begin{equation}
w_{v_1,v_2} = \exp\left( - \sum_{m=1}^{M} \theta_m^2 \cdot Q_{e^{(m)}}(v_1,v_2) \right),
\label{weight}
\end{equation}
where $\theta_m$ is the weight of feature $e^{(m)}$, and needs to be further learned to determine the AQI distribution of unlabeled nodes.

\vspace{-0.2cm}

\subsection{AQI Inference on Unlabeled Nodes}
The objective for the model's convergence is to minimize the model's uncertainty for inferring unlabeled nodes. We show that the distribution $p_u$ at an unlabeled node is the weighted average of distributions at its neighboring nodes~\cite{semi-supvised}. Then, the objective becomes to minimize the entropy of the whole model, i.e., $H(p_u) = -\sum_u p_u\log p_u$, to achieve accurate estimation.
This idea comes from the fact that an unlabeled node should possess a similar AQI value of its adjacent labeled nodes which are connected to it.
Therefore, based on the edge weight function in (\ref{weight}), we define the loss function of the correlation graph to enable the propagation between highly correlated nodes with higher edge weights:
\begin{equation}
\mathcal{L}(\bm{p}) = \sum_{(v_1,v_2)\in \mathcal{E}} \frac{1}{2} w_{v_1,v_2} \left\|p_{v_1}-p_{v_2}\right\|^2,
\end{equation}
where $p_{v_1}$ and $p_{v_2}$ are the AQI distribution at node $v_1$ and $v_2$, $\left\|p_{v_1}-p_{v_2}\right\| = \mathcal{D}_{KL}(p_{v_1}\| p_{v_2}) + \mathcal{D}_{KL}(p_{v_2}\| p_{v_1})$ denotes the similarity of AQI distributions between $p_{v_1}$ and $p_{v_2}$, described by the Symmetrical Kullback-Leibler~(KL) Divergence~\cite{deep learning book}. Thus, the objective function is given by:
\begin{equation}
\bm{p}^{*} = \mathop{\arg \min}_{\bm{p}} \ \mathcal{L}(\bm{p}).
\label{prob formula}
\end{equation}

By minimizing $\mathcal{L}(\bm{p})$, the nodes with higher edge weights would possess more similar AQI value while the nodes with lower edge weights would be more independent. Thus, the objective function can enable the AQI propagation between highly correlated nodes, thus improving inference accuracy.

\begin{proposition}
The solution of $p_u$ for (\ref{prob formula}) is the average of the distributions at its neighboring nodes.
\end{proposition}
\vspace{-0.1cm}
\begin{proof}
According to~\cite{semi-supvised}, the minimum function in (\ref{prob formula}) is \emph{harmonic}. Therefore, we have $\Delta p_u=0$ on unlabeled nodes $U$, while $\Delta p_l=P(v_l)$ on labeled nodes $L$. Here $\Delta$ is the $combinatorial$ $Laplacian$, which is defined by $\Delta = D-W$. $D=$~diag$(d_i)$ is the diagonal matrix with $d_i$ denotes the degree of $i$; $W=\{w_{i,j}\}$ is the weight matrix defined in (\ref{weight}).
The harmonic property provides the form of solution as:
\begin{equation}
p_u(x) = \frac{1}{d_u} \sum_{(u,l)\in \mathcal{E}} w_{u,l} p_l(x), \ \ x \in \left\{0,1,2,\dots,X \right\},
\end{equation}
where $X$ is the maximum possible AQI value. To normalize the solution, we redefine it as
\begin{equation}
\begin{split}
p_u(x) & = \frac{1}{d_u \sum_{x}p_u(x) } \sum_{(u,l)\in \mathcal{E}} w_{u,l} p_l(x) \\
& = \frac{\sum_{(u,l)\in \mathcal{E}}w_{u,l} p_l(x)}{\sum_{x} \sum_{(u,l)\in \mathcal{E}} w_{u,l} p_l(x) }.
\label{prob distribution}
\end{split}
\end{equation}
Hence, the distribution of unlabeled nodes $p_u$ is the average of distributions at its neighboring nodes.
\end{proof}

\begin{proposition}
$p_u$ in (\ref{prob distribution}) is a probability mass function~(PMF) on $\mathbf{x}$.
\label{proposition:pmf}
\end{proposition}
\vspace{-0.1cm}
\begin{proof}
To be a PMF on $\mathbf{x}$, we test the satisfaction of $p_u$ on the following three properties:
\begin{itemize}
\item[$\bullet$] The domain of $p_u$ is the set of all possible states of $\mathbf{x}$.
\item[$\bullet$] $\forall x\in \mathbf{x}$, $0\leq p_u(x)\leq 1$.
\item[$\bullet$] $\sum_{x\in \mathbf{x}} p_u(x) = 1$.
\end{itemize}

Considering the expression form in (\ref{prob distribution}), the conclusion is obvious, that $p_u$ is a PMF on $\mathbf{x}$.
\end{proof}
\vspace{0.2cm}

The solution again shows the influence of the highly correlated nodes that are connected by high-weight edges.

\subsection{Entropy-based Learning with AQI Scale Prior}
\noindent \textbf{AQI Scale Prior:} A key characteristic of our model is the conditioning of prior AQI scale knowledge on unlabeled nodes at current time stamp, $P(\cdot|\bm{\lambda})$~(see Fig.~\ref{fig:multi-layer graph model}). This conditioning allows the learnt AQI scale from vision-based sensing to guide ground WSN sensing, providing faster convergence and more accurate inference. Specifically, target space is divided into disjointed regions $\{R_1,R_2,\dots,R_k\}$ for aerial sensing. Each $R_j$ contains a number of cubes $\bm{C}^{(j)}$ to be inferred. For each $R_j$, the aerial sensing provides a conditioning $\lambda_j$ for $\bm{C}^{(j)}$:
\begin{equation}
\lambda_j: \left\{x_i \in \left[ X^{(j)}_{min},\ X^{(j)}_{max} \right],\quad \forall C_i \in \bm{C}^{(j)} \right\}.
\end{equation}
By applying $P(\cdot|\bm{\lambda})$ to $p_u$ in (\ref{prob distribution}), we finally induce $p_u(x|\bm{\lambda})$ as the inferred distribution. The conditioning brings faster convergence during training, and also enables more accurate inference. Sec. VI will detail the region division method, which helps lead out the low-cost wake-up mechanism design.

So far, the expression of $p_u$ is determined, the next step is to investigate the $learning$ weight functions given by (\ref{weight}). $\theta_m$ is learned from both labeled and unlabeled data, which forms a semi-supervised mechanism.

\noindent \textbf{Learning Criterion:} Since the labeled nodes are sparse, maximizing the likelihood of labeled nodes’ data to learn $\theta_m$ is infeasible. Instead, we use model’s entropy as the criterion, since high entropies can be regarded as unpredicted values, resulting in poor capability of inference and low accuracy. Thus, the objective is to minimize the entropy $H(p_u)$ of unlabeled nodes:
\begin{equation}
\min_{\bm{\theta}} \ H(p_u) \ = \ \min_{\bm{\theta}} \ \frac{1}{|U|} \sum_{i=1}^{|U|} H_i(p_i),
\end{equation}
where $|U|$ is the number of unlabeled nodes. By unfolding the objective function, we have
\begin{equation}
H(p_u) = - \sum_{i=1}^{|U|} \sum_{x=1}^{X} \frac{p_i\left(\mathbf{x}=x|\lambda_j\right)}{|U|}\log p_i\left(\mathbf{x}=x|\lambda_j\right).
\end{equation}
For simplicity, we denote $\sum_x p_j(\mathbf{x}=x|\lambda_j)\log p_j(\mathbf{x}=x|\lambda_j)$ as $p_j\log{p_j}$, the gradient can be derived as
\begin{equation}
\frac{\partial H}{\partial \theta_m} = \frac{1}{|U|} \sum_{j=1}^{|U|} \left( \log\frac{1}{p_j} -\frac{1}{\ln 2} \right) \frac{\partial p_j}{\partial \theta_m}.
\end{equation}
For every unlabeled $p_j$, we investigate $\frac{\partial p_j}{\partial \theta_m}$ based on (\ref{prob distribution}) and (\ref{weight}). By applying the chain rule of differentiation, the final gradient can be derived as
\begin{equation}
\frac{\partial w_{u,l}}{\partial \theta_m} = 2 w_{u,l} Q_{e^{(m)}} \theta_m.
\label{gradient}
\end{equation}

Thus, by iteratively learning and updating $\theta_m$ using (\ref{gradient}), the edge weights $\mathcal{W}$ can be studied and further generate the final AQI distribution when the iteration converges.

\begin{figure*}[!htbp]
\centering
\includegraphics[width=0.98\textwidth]{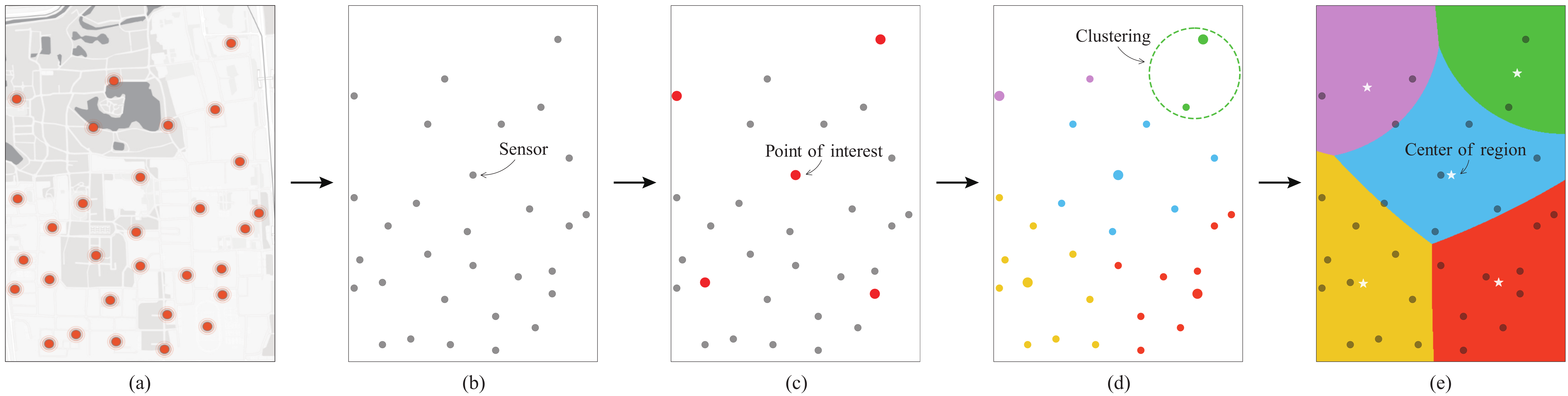}
\caption{An example of ground region division process: (a) ground sensors deployed in Peking University; (b) simplified graph illustrations; (c) dynamically select $k$ points of interests each time; (d) result of $k$-means clustering to different classes; (e) result of multi-site weighted voronoi division.}
\label{fig:region division}
\vspace{-0.2cm}
\end{figure*}

\noindent \textbf{Real-time Inference:} As illustrated in Fig.~\ref{fig:multi-layer graph model}, the real-time inference is based on (1) historical ground WSN data over last $d$ time stamps, and (2) the conditioning of prior AQI scale knowledge $P(\cdot|\bm{\lambda})$.
When the model converges, we obtain the determined AQI distribution $\hat{p}_u$ over $\mathcal{V}_U$, which is called as \emph{soft} labeling. To provide an exact or \emph{hard} labeling value of inference, as is proofed in Proposition~\ref{proposition:pmf} that $p_u$ is a PMF on $\mathbf{x}$, we quantize it using the expectation of $p_u$:
\begin{equation}
\widehat{P}_u = \mathbb{E}_{\mathbf{x}\sim p_u}\left[\mathbf{x} \right] = \sum_{x=1}^{X} x\cdot P_u\left(\mathbf{x}=x|\bm{\lambda}\right).
\label{hard labeling}
\end{equation}
Note that we can obtain $\widehat{P}_u(T_i)$ on each unlabeled node $u$ over $d$ time stamps. However, only data at current time $T_d$ is needed for real-time inference. Inspired by this idea, we store the whole inferred distribution map each time when real-time inference is completed, and further use it as historical data in the future sensing. By doing so, more labeled nodes $\mathcal{V}_L$ are known to get better inference results, which can accelerate the convergence speed and improve the accuracy.

\noindent \textbf{Future Forecasting:} Our model is also capable of future inference. In Fig.~\ref{fig:multi-layer graph model}, the edge can be extended to following time stamps and more. With the entropy-based learning procedure, it can maintain sufficient accuracy for near-future distribution forecasting even without the prior by aerial sensing.

\section{Energy-efficient Wake-up Mechanism}
Since our ground inference model is able to operate with very sparse labeled data, we only need to wake up a small number of ground sensors in selected regions to sense data at each $T_i$. This scheme can greatly save the battery of devices and extend our system's working duration, while also ensure high inference accuracy.

Recent methods~\cite{aircloud,mosaic,aqnet} which utilize ground WSN for inference, have employed all of their sensors to wake up simultaneously for data collection. This can lead to short working duration even with low-cost sensors. For example, devices in \cite{mosaic} can only last for less than 5 days before recharging, causing high consumption of battery power and human labor.
Yet there have scarcely been works in asynchronous wake-up for AQI monitoring. In fact, due to the spatial-temporal correlations of AQI distribution, waking up a specific number of sensors is enough to realize high inference accuracy, while greatly reducing the power consumption.
Thus, an energy-efficient wake-up mechanism is designed for ImgSensingNet to connect the aerial sensing and ground sensing, and to guide system selecting specific devices to wake up at current time stamp for energy saving.

\subsection{Voronoi Diagram based Region Division}
Since the total monitoring space can be very large, we first divide it into disjointed regions $\{R_1,R_2,\dots,R_k\}$ for aerial sensing. Note that even if devices are deployed in 3D~(e.g., different floors of buildings), we only consider 2D coordinates for region division. The height and the camera angle of UAV are fixed in advance, in order to make sure the region is covered in images. Cubes $\bm{C}^{(j)}$ inside each $R_j$ are provided an AQI scale conditioning $\lambda_j$ using vision-based inference. Since the distribution of ground devices is heterogeneous and uneven, we implement the division~(as shown in Fig.~\ref{fig:region division}) by following steps:

\noindent \textbf{Initialization:} Fig.~\ref{fig:region division}(a)(b)(c) present an example of the initialization process. Given a target space with ground devices deployed, $k$ points of interests~(POIs), e.g., a hospital or an office building, are selected dynamically at different time stamps.

\noindent \textbf{Clustering:} With $k$ POIs selected, we cluster each device to its nearby POI in spatial dimension based on the spatial correlation of AQI distribution, where $k$-means clustering is used. We obtain $k$ classes after the clustering, each containing $n_j$ devices~($j\in[1,k]$), as shown in Fig.~\ref{fig:region division}(d).

\noindent \textbf{Multi-site Weighted Voronoi Diagram:} Voronoi diagram is a partitioning of a plane into regions based on distance to sites in a specific subset~\cite{voronoi}. The original voronoi diagram only considers one site in a region, and using the Euclidean distance for division. As we have multiple devices in one region, we propose a \emph{multi-site weighted voronoi diagram} that enables division with (1) multiple sites inside one region, and (2) different weights assigned to each region for calculating the division boundary.

As shown in Fig.~\ref{fig:region division}(e), we first calculate the center $\phi_j$ in $R_j$ using the mean 2D coordinates of $n_j$ devices inside it. The coordinates of center $\phi_j$ is used for division on behalf of $R_j$. Since the number of devices $n_j$ varies over different regions, they should possess different weights when calculating the division boundary. Hence, we define the \emph{weighted distance} as:
\begin{equation}
D(y,\phi_j) = \frac{d(y,\phi_j)}{\sqrt{n_j}} = \frac{\left\| \vec{y} - \vec{\phi_j} \right\|_2}{\sqrt{n_j}},
\end{equation}
where $d(y,\phi_j)$ is the Euclidean distance between location $\vec{y}$ and region center $\vec{\phi_j}$, $n_j$ is the number of devices inside region $R_j$. Thus, the weighted voronoi division can be written as
\begin{equation}
V(\phi_i) = \bigcap_{j\neq i} \left\{ y \mid D(y,\phi_i) \leq D(y,\phi_j) \right\},\ \ i,j \in [1,k].
\end{equation}

\begin{proposition}
The complexity of the region division algorithm is $\Theta(k\cdot n)$.
\label{proposition:region divide}
\end{proposition}
\vspace{-0.1cm}
\begin{proof}
Denote the total number of 2D grids as $a\times b$, where $a$ and $b$ are constants for a specific monitoring area. In the first stage, all devices need to be clustered to a nearby POI, which computes for $k\cdot n$ times. In the second stage, we calculate the center of each class, which will compute $k$ times in the worst case. In the last stage, the assignments for each grid will take $a\cdot b\cdot k$ times for computing. Note that we always have $a\cdot b \ge n$, while there always exists a constant $c$ such that $c\cdot n \ge a\cdot b$~($c=a\cdot b$ in the worst case). Hence, the complexity for the last stage is $\Theta(a\cdot b\cdot k)=\Theta(k\cdot n)$. By combining three stages, we derive the final complexity as $\Theta(k\cdot n)+\mathcal{O}(k)+\Theta(k\cdot n)=\Theta(k\cdot n)$.
\end{proof}
\vspace{0.2cm}

\begin{algorithm}[!ht]
\caption{Voronoi Diagram based Region Division Algorithm}
\KwIn{current POIs $\{O_1,O_2,\dots,O_k\}$; device set $\bm{C}$}
\KwOut{regions $\{R_1,R_2,\dots,R_k\}$}\vspace{0.1mm}
\textbf{(a)} Initialize device set $\bm{C}^{(1)},\bm{C}^{(2)},\dots,\bm{C}^{(k)}$ in each region\;
\ForAll{$C_i \in \bm{C}$}{
    \textbf{(b)} Cluster $C_i$ to a nearest $O_j$, add $C_i$ into $\bm{C}^{(j)}$ \;
}
\ForAll{$\bm{C}^{(j)}$}{
    \textbf{(c)} Calculate the center $\phi_j$;
}
\textbf{(d)} Generate weighted voronoi diagram $\{R_1,R_2,\dots,R_k\}$.
\end{algorithm}

The best achieved complexity for classical voronoi diagram is $\mathcal{O}(n^2)$~\cite{voronoi}, which only fits for one-site division. In contrast, our algorithm can generate multi-site division as well as can reduce the computation overhead.
Algorithm 1 shows the procedure of the region division algorithm.

\subsection{When to Wake up}
At each time stamp, we first perform vision-based aerial sensing over $k$ regions to obtain the AQI scale inference for each region. Before triggering ground devices, we first utilize the semi-supervised learning model to give hard labeling on all nodes at current time stamp, based on the stored AQI inference maps over past $d$ time stamps. Therefore, for each node $C$, there are two estimations: (1) AQI scale $\left[ X_{min},\ X_{max} \right]$, and (2) pre-inferred value $\widetilde{X}$ using historical data. Based on the two priors, we propose an indicator to analyse the inference reliability and further decide which devices to wake up at current time stamp.

\begin{figure}[!htbp]
\centering
\includegraphics[width=0.4\textwidth]{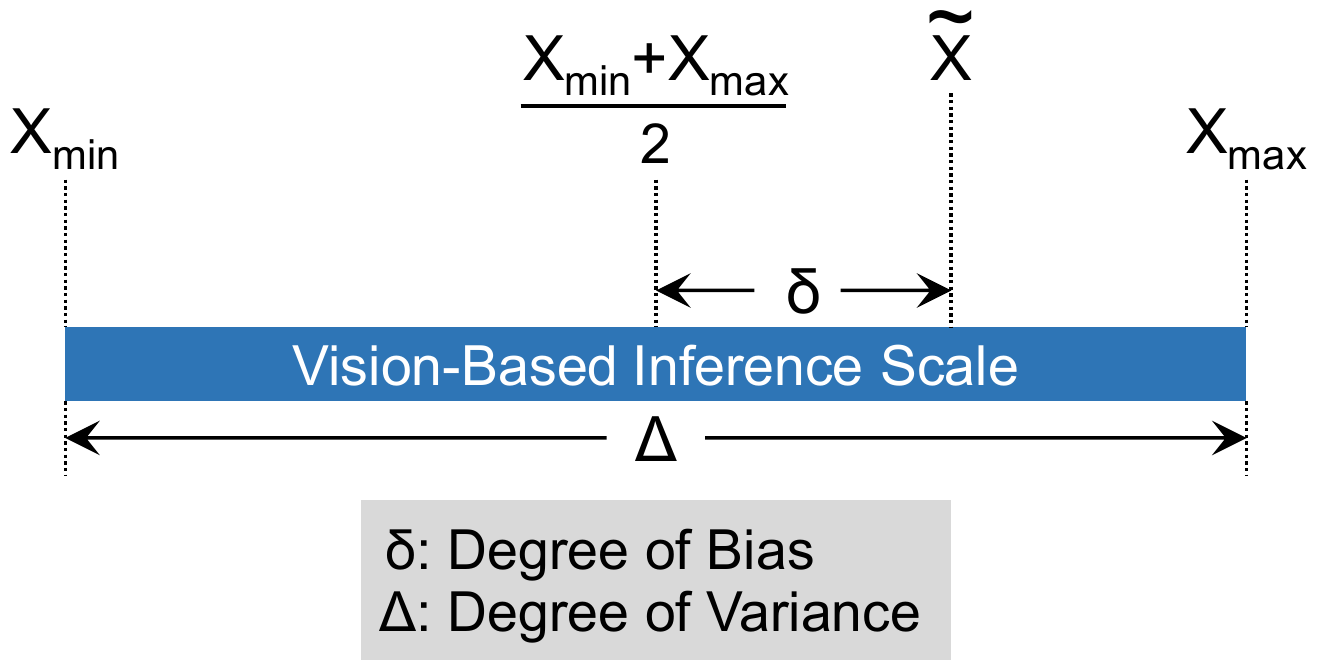}
\caption{An illustration of the two defined metrics, $\delta$ and $\Delta$.}
\label{fig:je}
\end{figure}

\noindent \textbf{Joint Estimation Error:} We first define two metrics of correlations between the two priors:
\begin{align}
\delta & = \left| \widetilde{X} - \frac{X_{min}+X_{max}}{2} \right|, \ \ \ \ (DoB) \\
\Delta & = X_{max} - X_{min}, \qquad \qquad \ \ (DoV)
\end{align}
where we call $\delta$ as \emph{Degree of Bios}~(DoB), $\Delta$ as \emph{Degree of Variance}~(DoV), as shown in Fig.~\ref{fig:je}. Intuitively, when $\Delta$ is low, the variance of the AQI scale prior is small, which means a more reliable inference; as for $\delta$, a low $\delta$ induces small deviation between the two priors, which in turn guarantees the inference reliability. Hence, DoB and DoV can both reflect the degree of estimation errors. By merging the two metrics, we define the \emph{Joint Estimation Error}~(JE) as:
\begin{equation}
JE = \frac{1}{2} \left[ \frac{\delta}{\delta_m} + \frac{\Delta}{\Delta_m} \right],
\end{equation}
where $\delta_m$ and $\Delta_m$ denote the maximum value of DoB and DoV for all nodes with devices. As a result, \emph{JE} is normalized into $[0,1]$, and each node has a corresponding \emph{JE}. In general, \emph{JE} reflects the degree of average inference error for labeled nodes before waking up for ground sensing.
For $i^{th}$ cube, a greater $\text{\emph{JE}}^{(i)}$ indicates higher uncertainty for inference at $C_i$, which signifies $C_i$ should be measured currently if $\text{\emph{JE}}^{(i)}$ exceeds a threshold. Hence, given a specific \emph{JE} as threshold, sensors/nodes with $\text{\emph{JE}}^{(i)}\ge\text{\emph{JE}}$ should wake up for data collection at current time stamp.
These nodes are then labeled with measured data at layer $T_d$, which can best reduce the model's entropy and are sufficient for real-time and future inference. In this way, by only measuring a small number of cubes, ImgSensingNet can greatly reduce the measurement overhead while maintaining high inference accuracy.

In general, \emph{JE} is adjusted manually for different scenarios, which forms a tradeoff. When \emph{JE} is low, the threshold for inference error declines, indicating the measuring cubes will increase and brings promotion in inference accuracy. However, it can cause great battery consumption. On the other hand, as \emph{JE} is high, the measuring cubes will decrease. This may cause a decline in accuracy, but can highly reduce consumption. In summary, the tradeoff between accuracy and consumption should be studied to acquire a better performance.

\begin{figure}[!htbp]
\centering
\includegraphics[width=0.48\textwidth]{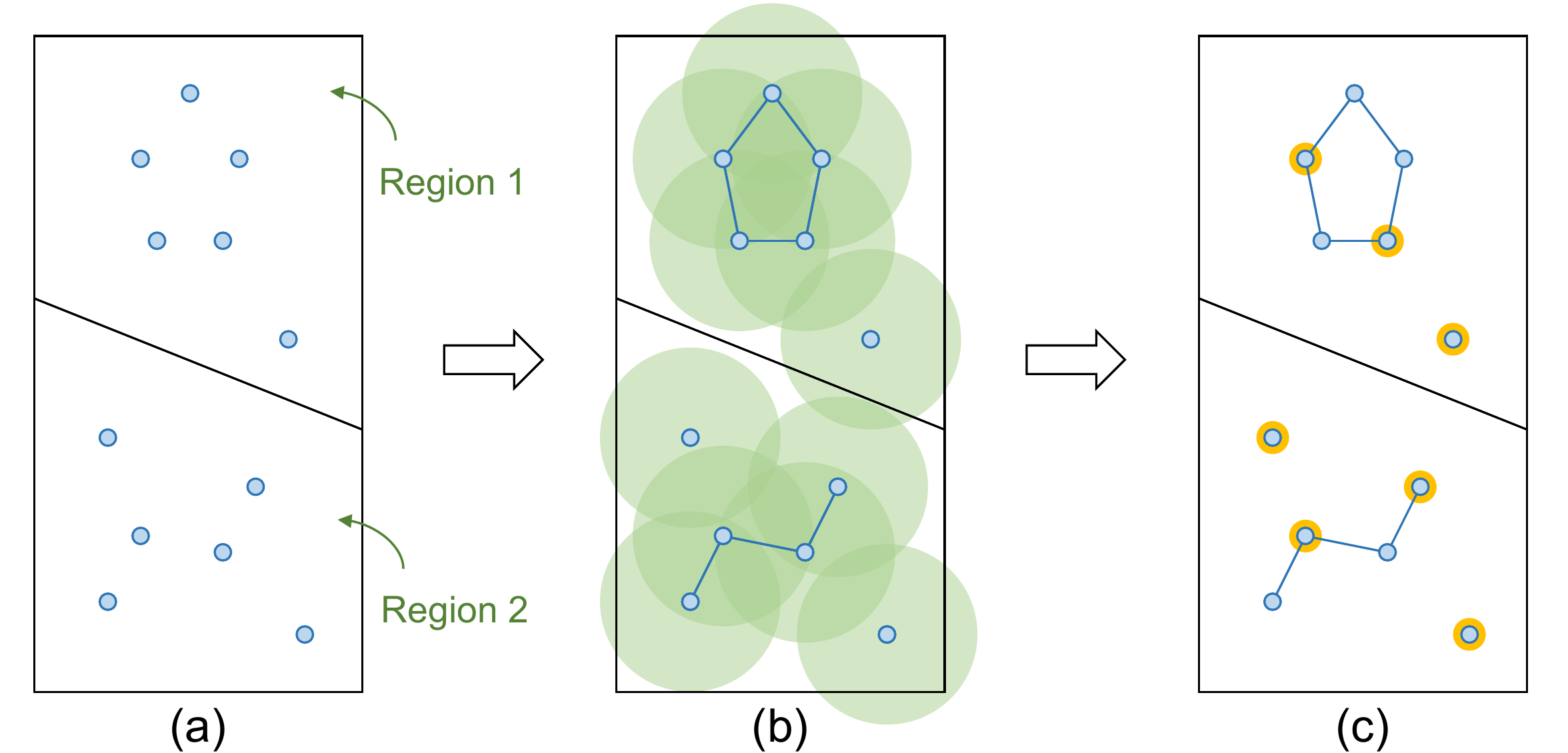}
\caption{An example of the final wake-up set $\mathcal{S}$ construction: (a) the selected wake-up nodes set $\mathcal{M}$ based on \emph{JE}; (b) the spatial neighbors within radius $r$ in graph model; (c) finding the min independent dominating set $\mathcal{S}$.}
\label{fig:min dominating set}
\end{figure}

\noindent \textbf{Wake-up Mechanism Design:} \emph{JE} can guide system waking up selected devices at each time stamp. First, if the two priors $\left[ X_{min},\ X_{max} \right]$ and $\widetilde{X}$ are both less than a pre-given value $\sigma$, then the current air quality is too good to wake up the device for measurement, and switching off the device in such a case can help save the battery.

Second, since we construct the graph model by connecting nodes within a spatial radius $r$, it is possible that two nodes that are selected to wake-up are adjacent and connected in the model~(Fig.~\ref{fig:min dominating set}(a) provides an example). In this case, waking up connected nodes would be redundant as their measurements are similar. Denote $\mathcal{M}$ as the set of selected wake-up nodes using \emph{JE}, our objective is to find a subset $\mathcal{S}\subseteq \mathcal{M}$ such that (1) nodes in $\mathcal{S}$ are not adjacent, and (2) every node not in $\mathcal{S}$ is adjacent to at least one member of $\mathcal{S}$. This problem is well-known as the \emph{minimum independent dominating set} problem~\cite{dominating set}, which is NP-hard. Since $\mathcal{M}$ is sparse in our case, the computing overhead is small, we simply apply a greedy-based method to find $\mathcal{S}$.
Note that the algorithm is applied in each region independently. 
The total process of finding a final wake-up set $\mathcal{S}$ based on $\mathcal{M}$ is shown in Fig.~\ref{fig:min dominating set}.

\begin{algorithm}[!ht]
\caption{Wake-up Mechanism Design}
\KwIn{device set $\bm{C}^{(1)},\bm{C}^{(2)},\dots,\bm{C}^{(k)}$; \emph{JE}; $\sigma$}
\KwOut{target node set $\{\mathcal{S}\}$}
\ForAll{$\bm{C}^{(j)}$}{
\ForAll{$C^{(j)}_i\in \bm{C}^{(j)}$}{
    \textbf{(a)} Compute $\emph{\text{JE}}^{(i)}$, $\widetilde{X}^{(j)}$, $( X_{min}^{(j)},\ X_{max}^{(j)} )$\;\vspace{0.3mm}
    \textbf{(b)} Initialize $\mathcal{M}^{(j)} \gets \varnothing$\;\vspace{0.2mm}
\If{$\underline{\textit{JE}^{(i)}\ge \textit{JE}}$ \textbf{and} $\underline{\max{\{\widetilde{X}^{(j)}, X_{max}^{(j)}\}}\leq \sigma}$}{\vspace{0.1mm}
    \textbf{(c)} Add $C^{(j)}_i$ into $\mathcal{M}^{(j)}$ \;
}}
\While{$\mathcal{M}^{(j)} \neq \varnothing$}{
\textbf{(d)} Choose node $S\in \mathcal{M}^{(j)}$ with max $d_s$, add $S$ into $\mathcal{S}^{(j)}$\;
\textbf{(e)} Remove $S$ and all it's neighbors from $\mathcal{M}^{(j)}$\;
}
}
$\{\mathcal{S}\} = \bigcup \left\{ \mathcal{S}^{(1)},\mathcal{S}^{(2)},\dots,\mathcal{S}^{(k)}\right\}$.
\end{algorithm}

\begin{lemma}
The number of final wake-up devices $|\mathcal{S}|$ decreases monotonically when $r$ increases.
\label{lemma:wakeup mechanism}
\end{lemma}
\vspace{-0.1cm}
\begin{proof}
When $r$ increases to $r^{'}$, the number of edge $\mathcal{E}^{'}$ can increase, which induces larger local connectivity graphs. For the extreme case, we choose $\mathcal{S}^{'}$ as the same as $\mathcal{S}$, which at least forms an independent dominating set. Since $\mathcal{E}^{'}$ can increase, nodes in $\mathcal{S}$ can be directly connected, which cannot be a minimum set. Hence, we have $|\mathcal{S}^{'}|\leq |\mathcal{S}|$ with $r^{'}$, that $|\mathcal{S}|$ decreases monotonically when $r$ gets larger.
\end{proof}

\begin{proposition}
The complexity of the wake-up mechanism algorithm decreases monotonically when $r$ increases, which is $\Theta(|\mathcal{S}|\cdot n)$.
\label{proposition:wakeup mechanism}
\end{proposition}
\vspace{-0.1cm}
\begin{proof}
For the inner loop, we compute $n_j$ times to generate $\mathcal{M}^{(j)}$ for each region $R_j$. As we need to traverse $\mathcal{M}^{(j)}$ to find a node $S$ with largest degree $d_s$ each time, and the total times are represented by $|\mathcal{S}^{(j)}|$, thus the complexity to find $\mathcal{S}^{(j)}\subseteq \mathcal{M}^{(j)}$ is $\mathcal{O}(|\mathcal{S}^{(j)}|\cdot n_j)$.
As for the outer loop, $k$ times are needed for each region. Hence, the total complexity is $\mathcal{O}(k\cdot(|\mathcal{S}^{(j)}|\cdot n_j + |\mathcal{S}^{(j)}|)) = \mathcal{O}(k\cdot |\mathcal{S}^{(j)}|\cdot n_j)$. It's obvious to find both the upper bound and lower bound can be denoted as $|\mathcal{S}|\cdot n$ multiply by a constant $c$, which induces the final complexity as $\Theta(|\mathcal{S}|\cdot n)$.

Based on Lemma~\ref{lemma:wakeup mechanism}, the complexity also decreases monotonically when $r$ gets larger.
\end{proof}
\vspace{0.2cm}

Algorithm 2 shows the procedure of the wake-up mechanism. The target node set is determined based on \emph{JE} as well as two conditions we studied, which contains the nodes with the highest uncertainty currently.
Thus, by only monitoring $\mathcal{S}$, ImgSensingNet achieves high inference accuracy while greatly reducing the measurement overhead.

\begin{proposition}
The overall complexity for wake-up mechanism each time is $\Theta\left((|\mathcal{S}|+k)\cdot n\right)$.
\end{proposition}
\vspace{-0.1cm}
\begin{proof}
The proposition is obvious based on the results of Proposition~\ref{proposition:region divide} and \ref{proposition:wakeup mechanism}.
\end{proof}
\vspace{0.2cm}

The overall wake-up mechanism contains dynamic region division and wake-up devices selection each time. In practice, since $k$ and $|\mathcal{S}|$ are small due to the device sparsity in deployment, by choosing proper $r$, the computation of the whole mechanism can be completed in real-time, which will be shown in the evaluation.

\section{Implementation}
In this section, we detail the implementation of ImgSensingNet system. Specifically, we first introduce the components of ImgSensingNet, including both the ground devices and the aerial device. Based on the devices, we then illustrate the data collection of both aerial image data and ground sensing data.

\begin{figure}[!t]
\centering
\includegraphics[width=0.48\textwidth]{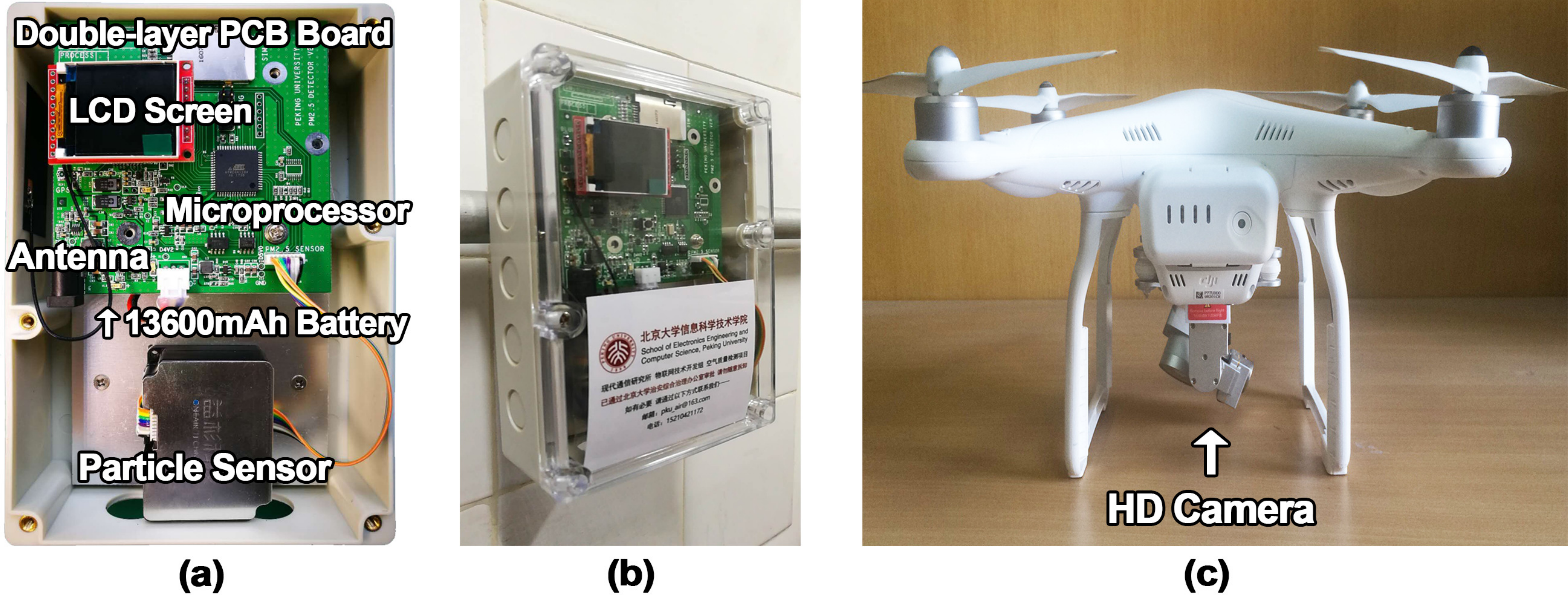}
\caption{The components of ImgSensingNet: (a) the sensor-based monitoring device; (b) the device is installed on the wall; (c) the UAV for vision-based sensing.}
\label{fig:hardware}
\end{figure}

\subsection{System Components}
\textbf{Ground Devices:} Fig.~\ref{fig:hardware}(a) shows the components of the ground AQI monitoring device. Each device contains a low-cost \texttt{A3-IG} sensor, a two-layer circuit board, a \texttt{ATmega128A} working as the micro-controller unit~(MCU), a \texttt{SIM7000C} as the wireless communication module, a 13600mAh rechargeable battery and a fixed shell structure.
Considering the intrinsic lack of precision for the small-scale laser-based sensors, these devices are carefully calibrated through a whole month adjustment by comparing the results with a high-precision calibrating instrument \texttt{TSI8530}.
Finally, these devices can provide $\leq 3\%$ monitor error for common pollutants in AQI calculation, such as ${\rm PM}_{2.5}$ and ${\rm PM}_{10}$, and send the real-time data back to the central server for further data analysis.
To realize high energy-efficiency, the devices are programmed to sleep during most of the time and wake up for data collection based on adjustable time intervals that are controlled by a designed wake-up mechanism, which will be discussed in Sec. VI. Thus, an online tradeoff between data quantity and battery endurance can be implemented.

\textbf{Aerial Device:} For the UAV, we select \texttt{DJI} \texttt{Phantom} \texttt{3} Quadcopter as the sensing device, as shown in Fig.~\ref{fig:hardware}(c). The GPS sensor on the UAV can provide the real-time 3D position. In existing systems~\cite{IoTJ}, the UAV can keep flying for at most 10$\sim$20 minutes due to both the \emph{load consumption}~(carrying sensors can significantly reduce the UAV's battery life), and the \emph{loitering consumption}~(to acquire sensing data, the UAV needs to stay still at every measuring location), which restricts the monitoring scope within one measurement~\cite{optimal}.
However, as the UAV contains a built-in HD camera, when we focus on vision-based sensing, the extra loading and the hovering time can be eliminated. Hence, the sensing scope as well as the flight duration can be greatly increased. 

\subsection{Experiment Setup and Data Collection}

\begin{figure}[!htbp]
\centering
\includegraphics[width=0.48\textwidth]{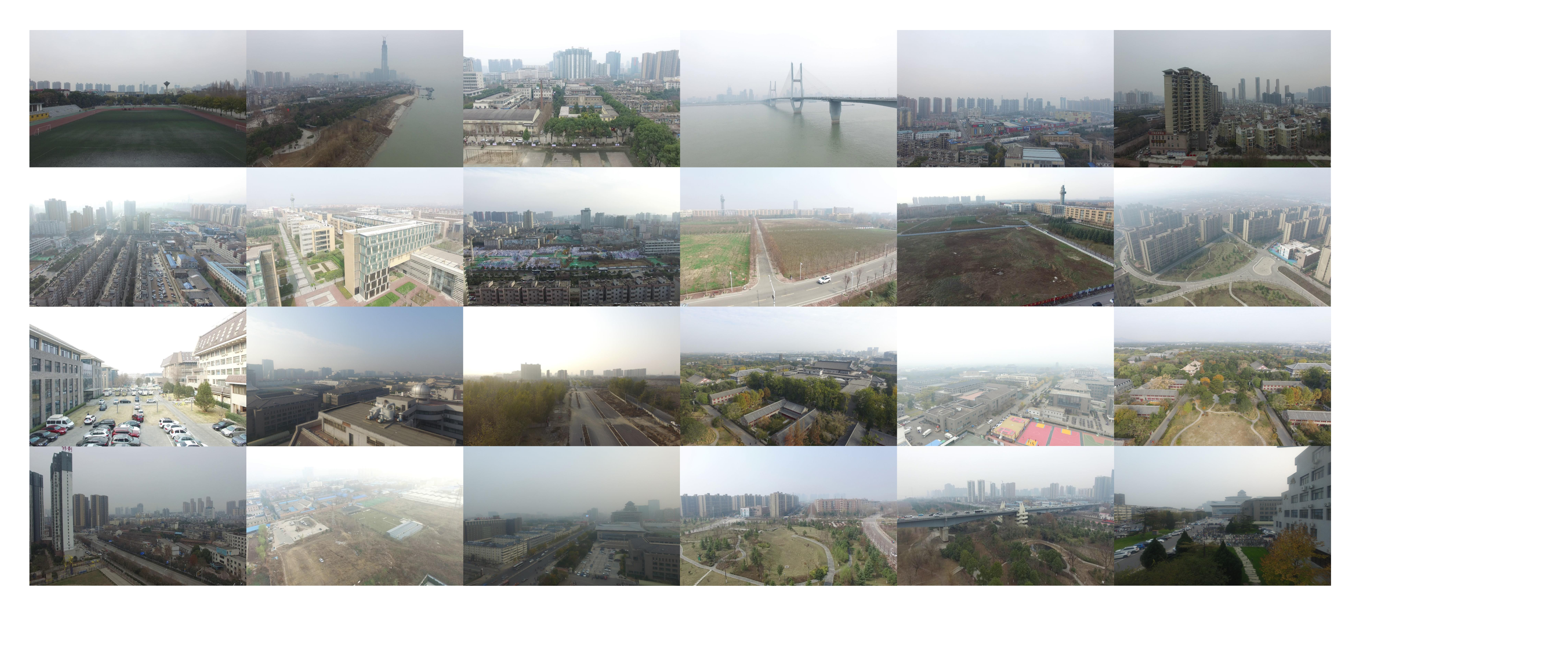}
\caption{An overview of the aerial image dataset: more than 10,000 labeled haze images taken in Beijing, Xi'an and Wuhan.}
\label{fig:img dataset}
\end{figure}

\begin{table*}[!htbp]
\caption{Average Estimation Errors by Different Methods}
\centering
\begin{tabular}{ c|c|c|c|c|c|c|c|c }
\hline
\multicolumn{1}{ c|}{\multirow{2}*{Methods}} & \multicolumn{4}{|c}{2D} & \multicolumn{4}{|c}{3D} \\
\cline{2-9}
{} & \emph{Real-time} & \emph{After 1 hour} & \emph{After 3 hours} & \emph{After 10 hours} & \emph{Real-time} & \emph{After 1 hour} & \emph{After 3 hours} & \emph{After 10 hours} \\
\hline
\textbf{ImgSensingNet} & \textbf{3.540} & \textbf{6.178} & \textbf{9.330} & \textbf{20.269} & \textbf{5.529} & \textbf{9.928} & \textbf{13.341} & \textbf{31.409} \\
\hline
ARMS~\cite{IoTJ} & 5.412 & $-$ & $-$ & $-$ & 7.384 & $-$ & $-$ & $-$ \\
\hline
LSTM Nets~\cite{lstm} & 4.217 & 7.804 & 10.672 & \textbf{19.873} & $-$ & $-$ & $-$ & $-$ \\
\hline
AQNet~\cite{aqnet} & 4.493 & 7.562 & 13.695 & 25.192 & 6.481 & 14.735 & 19.634 & 39.790 \\
\hline
S-T $k$NN~\cite{STkNN} & 7.039 & 9.667 & 11.882 & 26.055 & 9.147 & 12.954 & 18.065 & 44.256 \\
\hline
\end{tabular}
\label{table:eval accuracy}
\vspace{-0.3cm}
\end{table*}

\begin{figure}[!t]
\centering
\subfigure{
	\label{fig:3d cnn compare acc}
	\includegraphics[width=0.236\textwidth]{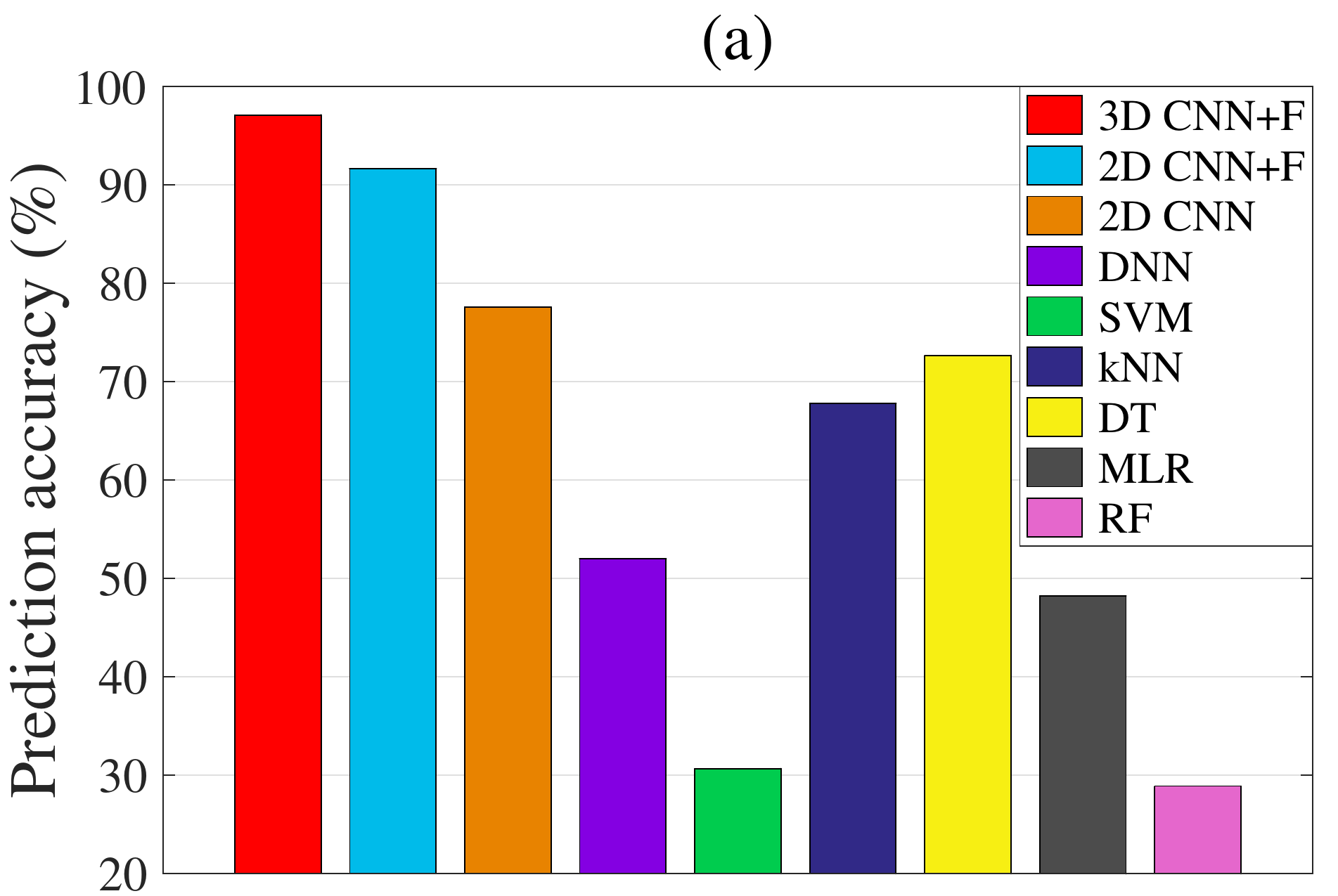}
}
\hspace{-2.6ex}
\subfigure{
	\label{fig:3d cnn compare rmse}
	\includegraphics[width=0.236\textwidth]{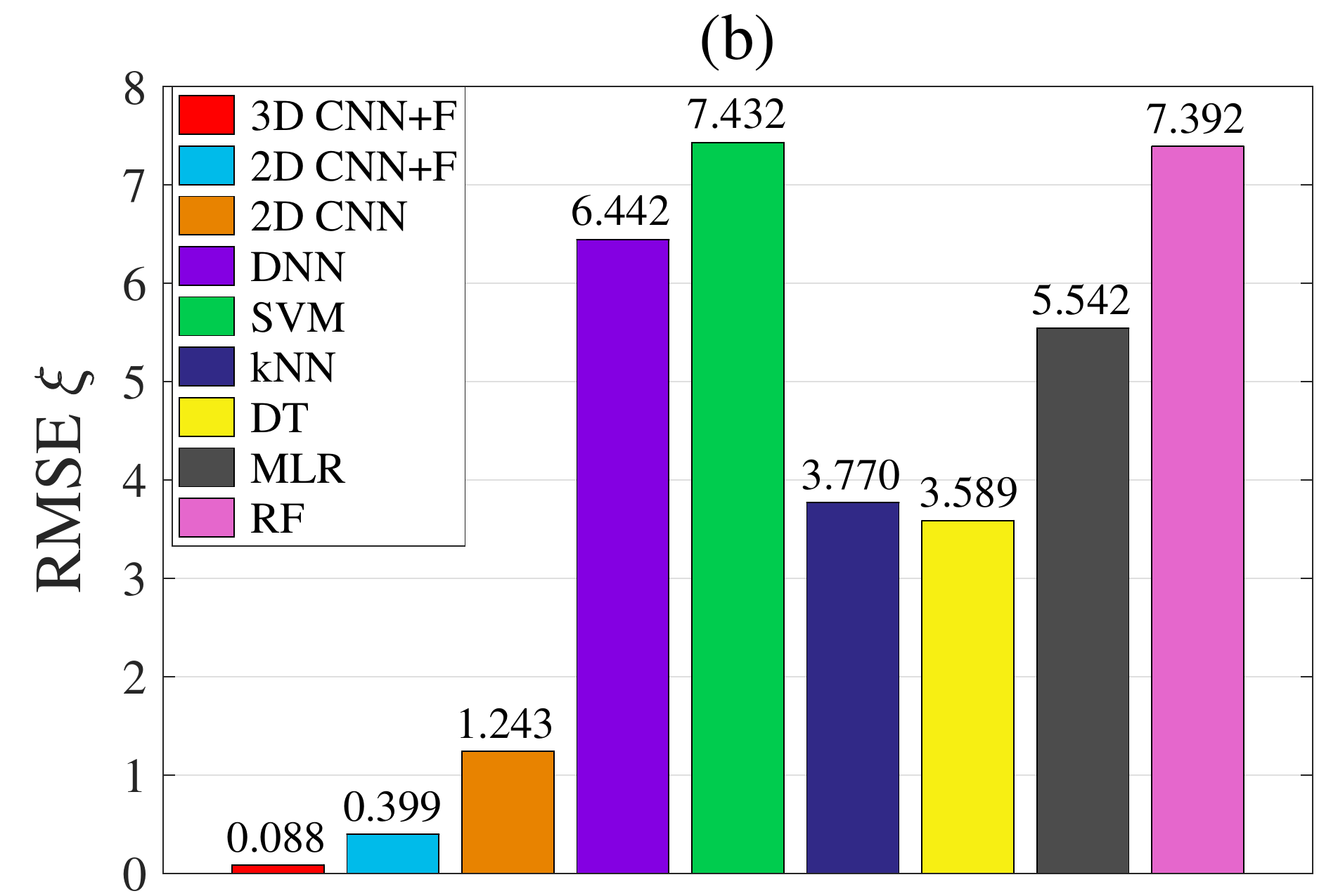}
}
\caption{The image-based inference comparison between different methods: (a) the inference accuracy; (b) the robustness of inference.}
\label{fig:3d cnn compare}
\vspace{-0.3cm}
\end{figure}

ImgSensingNet system prototype includes 200 ground devices and a UAV, and it has been deployed on two university campuses~(i.e., Peking University and Xidian University), since Feb. 2018.
Throughout more than a half year’s measurement, 2.6 millions of ground data samples and and a number of 17,630 haze images are collected, covering from good air quality cases to hazardous air quality cases, which are used for evaluation in this paper.

\textbf{Aerial Image Data:} The vision-based sensing works online continuously and real-timely by sampling images from the UAV video streams between equal time intervals. To get ground truth data for training the CNN model, we set up the dataset by carrying calibrated sensor to label the image with ground truth AQI value. We collected 17,630 labeled images in different places to make the data generalize well.
Fig.~\ref{fig:img dataset} shows an overview of the image dataset.

\textbf{Ground Sensing Data:} The testing areas are on campus of Peking University~($\sim$ 2km$\times$2km) and Xidian University~($\sim$ 2km$\times$1.5km). The ground devices are deployed in 3D space with a 50m maximum height. We divide the areas into 20m$\times$20m$\times$10m cubes, where a small number of cubes are deployed with our devices. We manually set the minimum time intervals as 30 minutes.

\section{Evaluation}
In this section, we present the performance analysis of ImgSensingNet in various aspects.

\subsection{Vision-based Aerial Sensing}
We evaluate the proposed AQI scale inference model in two aspects: \emph{accuracy} and \emph{robustness} in predictions. We randomly divide the image dataset with 7:3 training set to testing set ratio. We compare the proposed inference model with the following models from two categories: (1) three \emph{deep learning} methods: 2D CNN with our extracted features, 2D CNN without features and a 50-layer deep neural network~(DNN); the 2D CNN architecture is the same with our 3D CNN, but with only 2D kernels. (2) five \emph{classical training} methods: support vector machine~(SVM), $k$-nearest neighbors~($k$NN), decision tree~(DT), multi-variable linear regression~(MLR) and random forest~(RF).

\textbf{Accuracy of Inference:} As shown in Fig.~\ref{fig:3d cnn compare acc}, in general our method outperforms all other models. We can achieve a 96\% accuracy for image-based AQI scale inference by the proposed model. Moreover, when the features are considered, the 2D CNN model also outperforms the one without features, which confirms the effectiveness of haze-relevant feature extraction.

\textbf{Robustness of Inference:} In Fig.~\ref{fig:3d cnn compare rmse}, we test how much the inferred values deviate from the real values, using root mean square error~(RMSE). The results show that the proposed model outperforms other models by maintaining a very low deviation, i.e., 0.088 classification deviation in average. This again proves the advances in using 3D model and feature extraction.

\begin{figure}[!t]
\centering
\subfigure{
	\label{fig:consume compare uav}
	\includegraphics[width=0.24\textwidth]{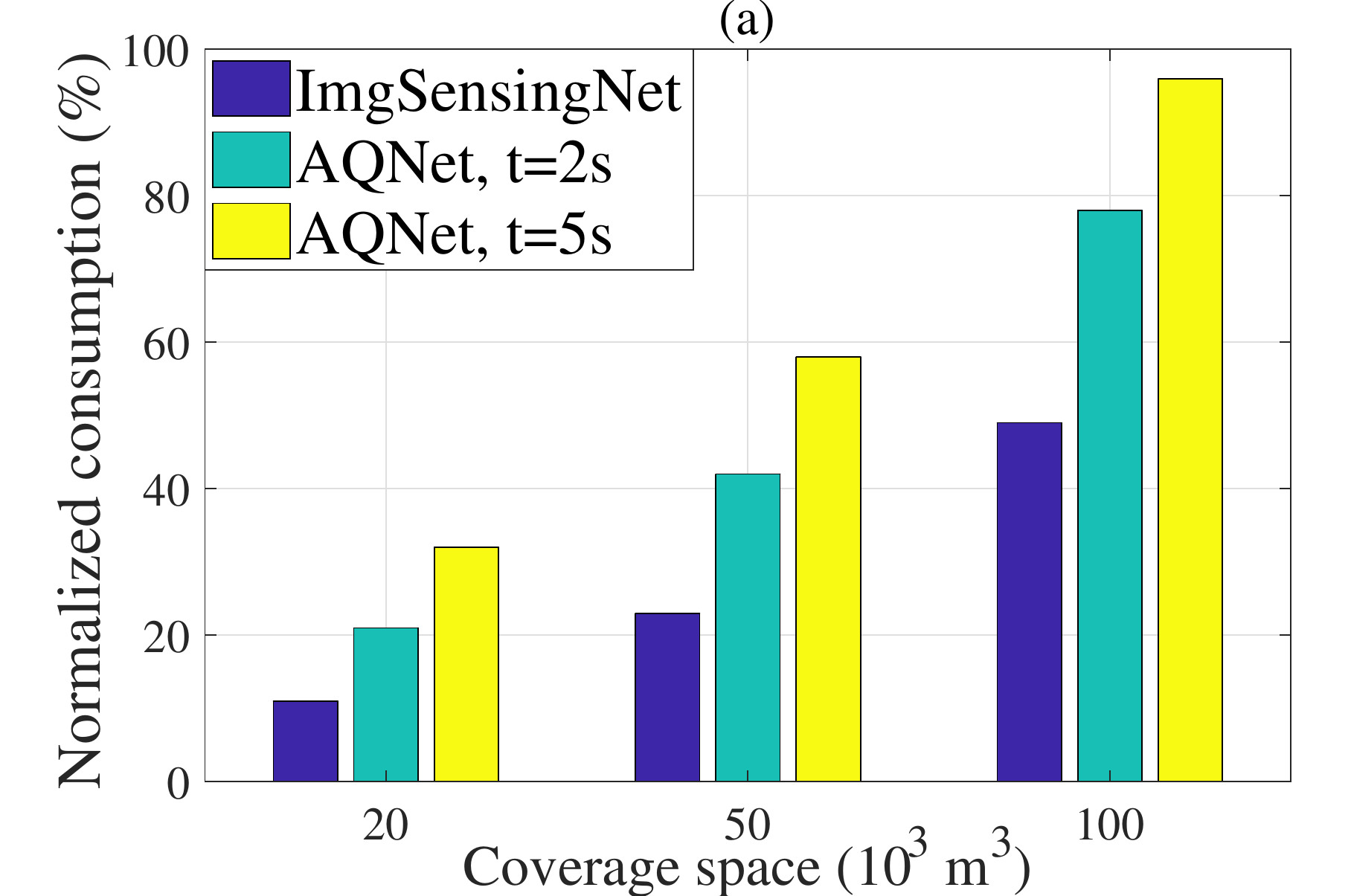}
}
\hspace{-3.6ex}
\subfigure{
	\label{fig:consume compare wsn}
	\includegraphics[width=0.24\textwidth]{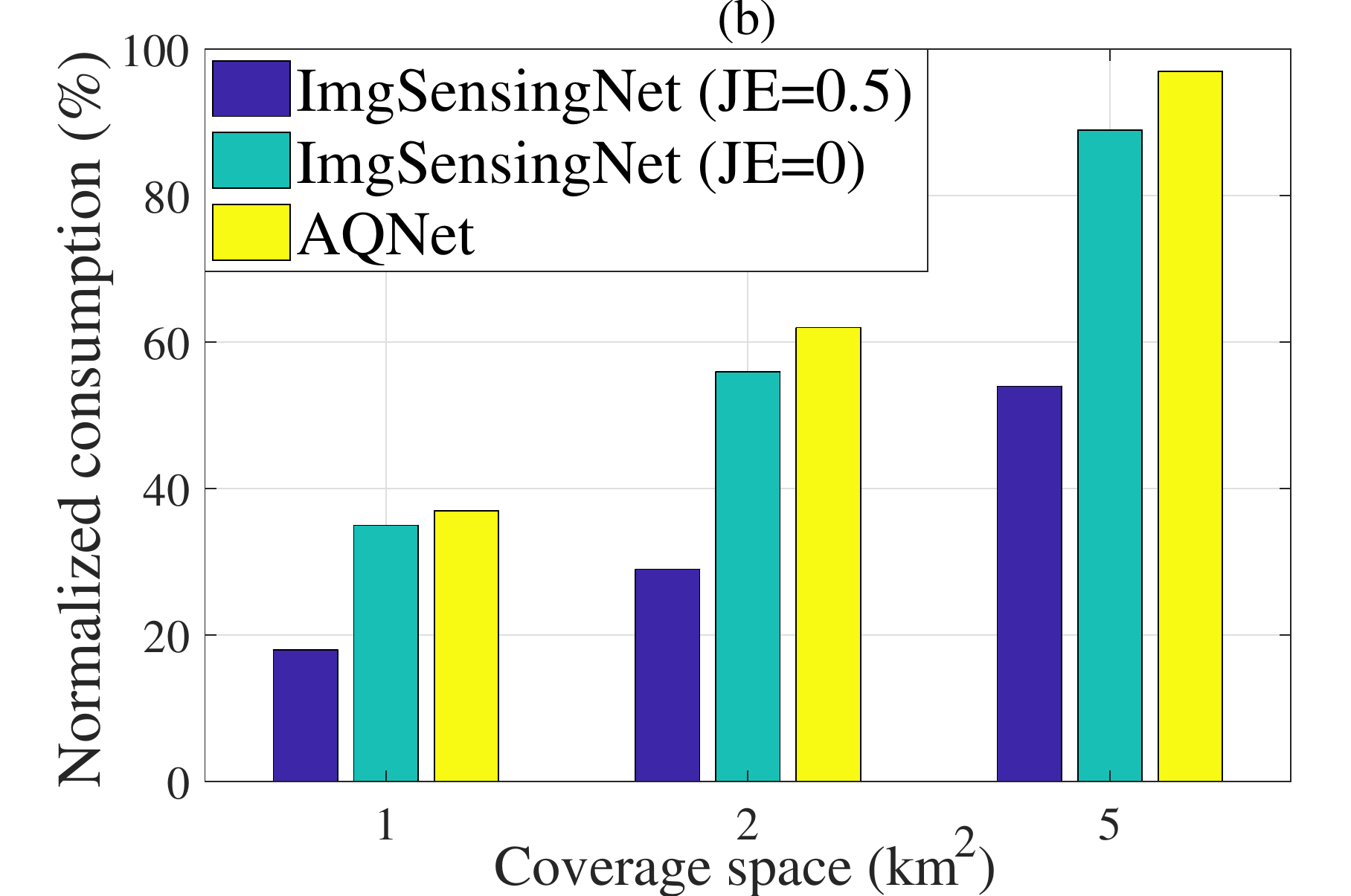}
}
\caption{The energy-efficiency comparison between different methods: (a) the consumption of aerial UAV sensing; (b) the consumption of ground WSN sensing.}
\label{fig:consume compare}
\vspace{-0.3cm}
\end{figure}

\begin{figure*}[!t]
\centering
\subfigure{
    \label{fig:wakeup device 30}
    \includegraphics[width=0.22\textwidth]{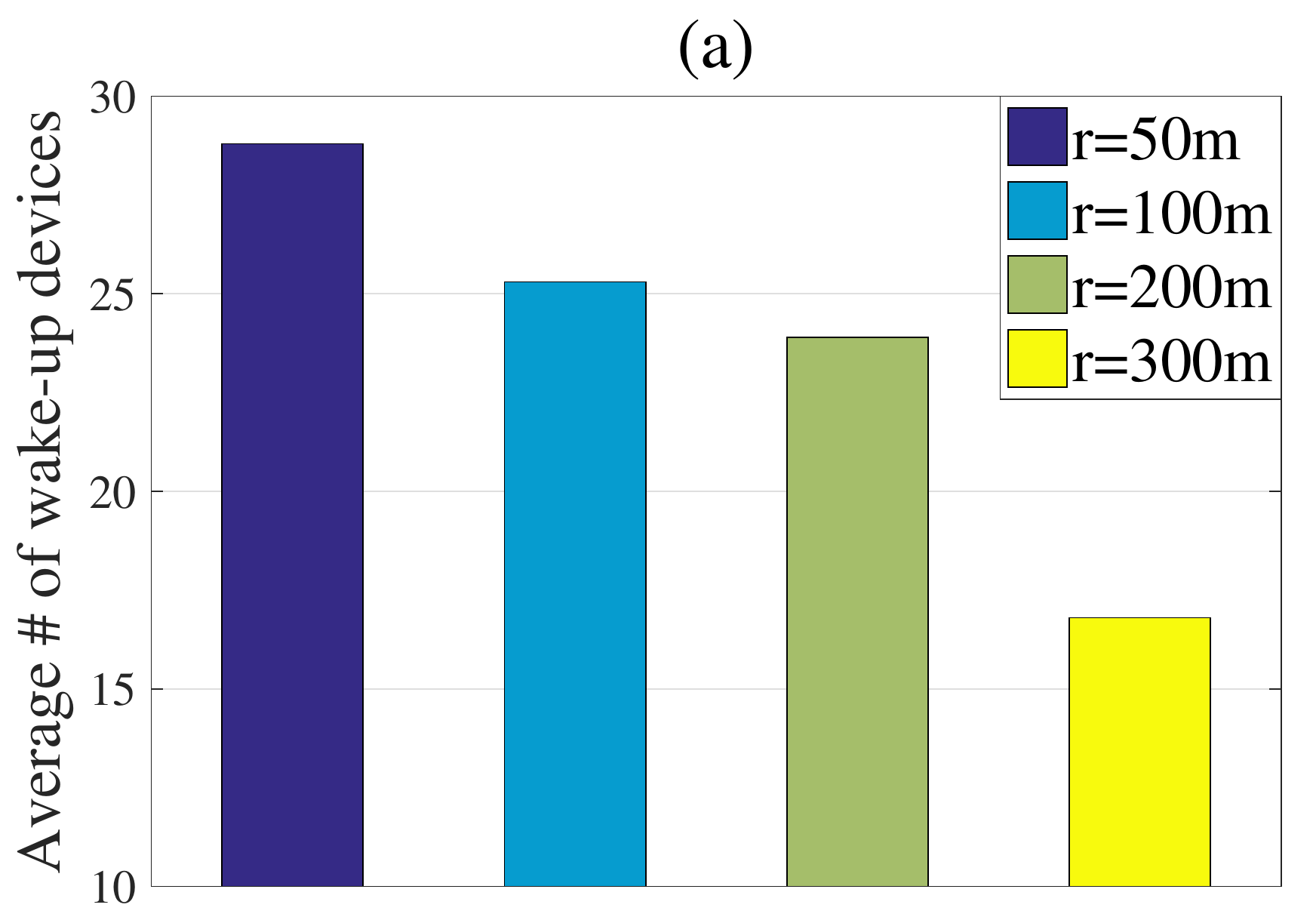}
}
\hspace{1ex}
\subfigure{
    \label{fig:wakeup runtime 30}
    \includegraphics[width=0.22\textwidth]{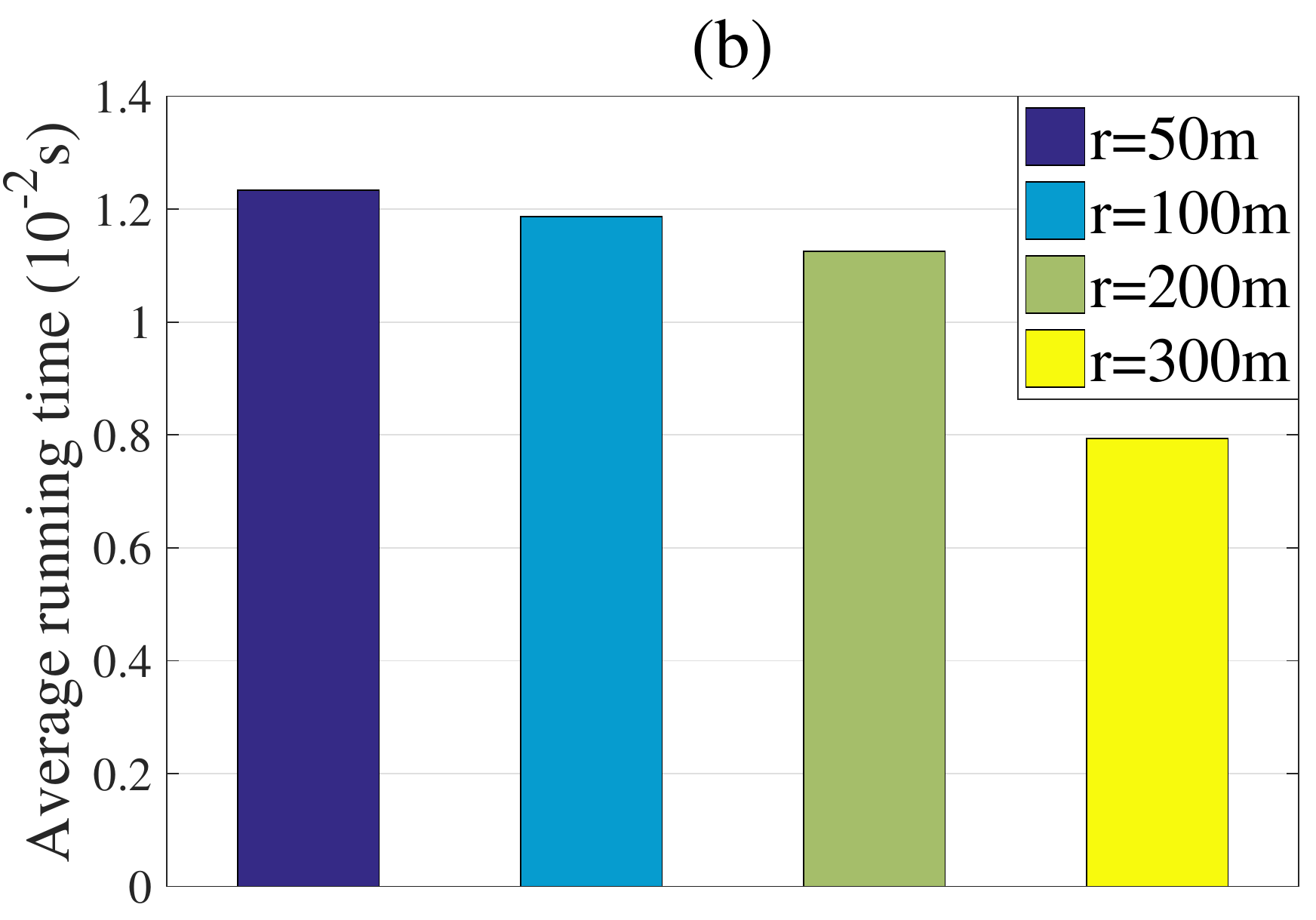}
}
\hspace{1ex}
\subfigure{
    \label{fig:wakeup device 100}
    \includegraphics[width=0.22\textwidth]{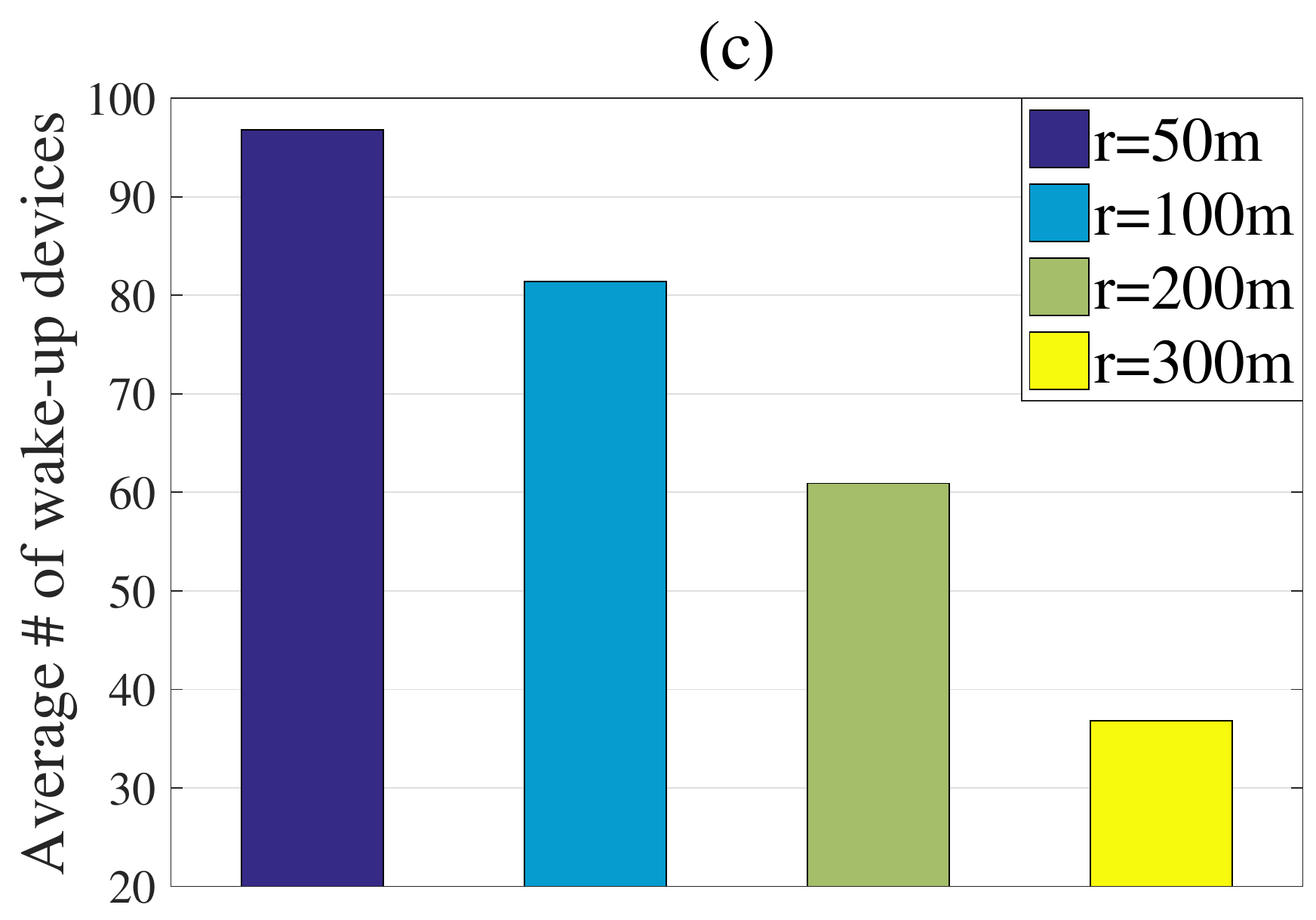}
}
\hspace{1ex}
\subfigure{
    \label{fig:wakeup runtime 100}
    \includegraphics[width=0.22\textwidth]{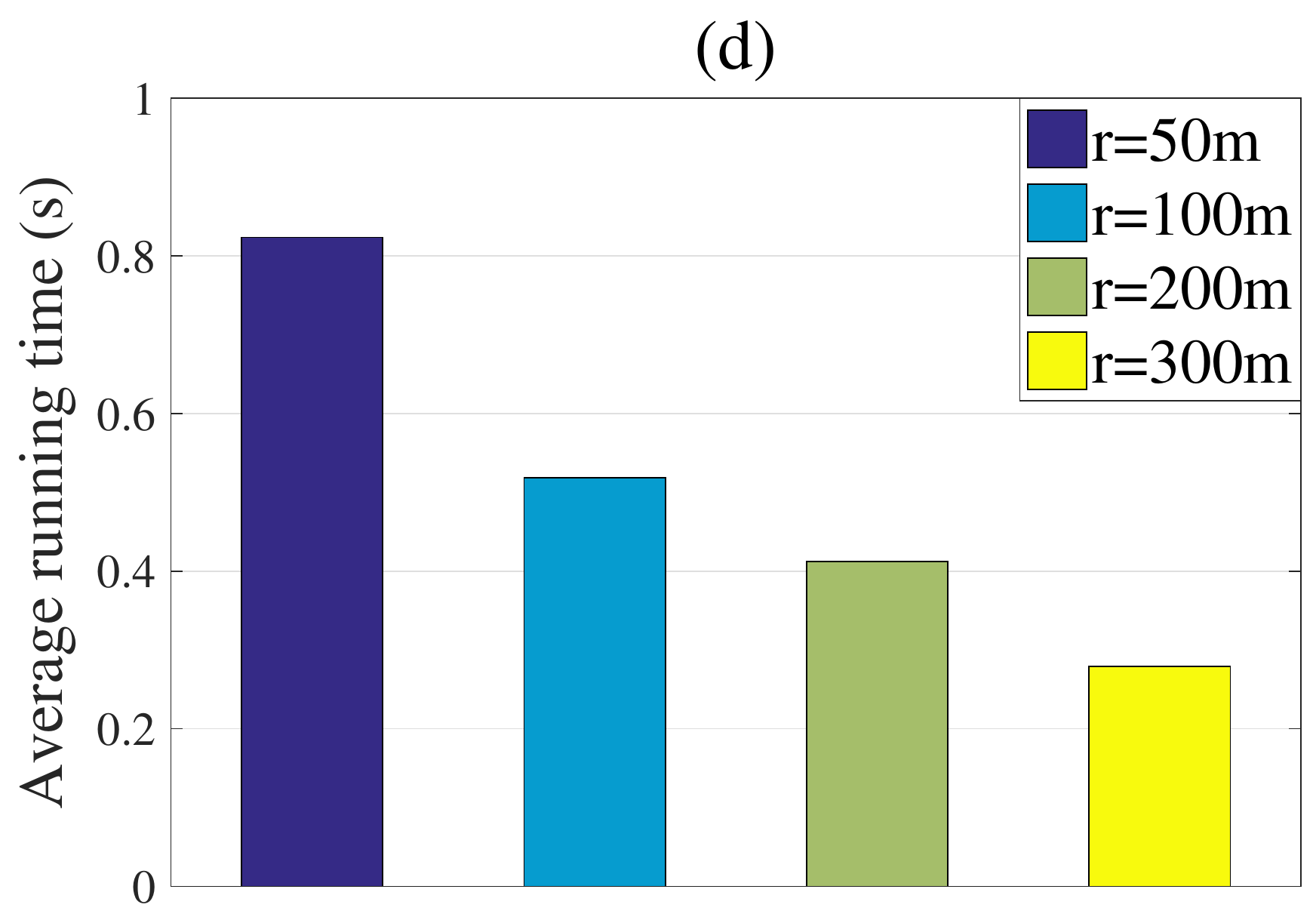}
}
\caption{The wake-up mechanism performance versus different $r$: (a) the average number of wake-up devices, with 30 total devices; (b) the average runtime, with 30 total devices; (c) the average number of wake-up devices, with 100 total devices; (d) the average runtime, with 100 total devices.}
\label{fig:wakeup}
\vspace{-0.3cm}
\end{figure*}

\subsection{Inference Accuracy}
We evaluate the inference accuracy of ImgSensingNet in both real-time estimation and near-future forecasting.
Since there are no measured data for most cubes, we divide labeled samples into training set and testing set, while performing an cross-validation by randomly choosing the training data, and repeat for 1000 times to avoid stochastic errors.

We use inference models in state-of-the-art AQI monitoring systems as ARMS~\cite{IoTJ}, LSTM Nets~\cite{lstm}, AQNet~\cite{aqnet} and spatio-temporal $k$NN~\cite{STkNN} for comparison. These models are all evaluated using the same data each time.

In Table~\ref{table:eval accuracy}, we report the average estimation errors of real-time inference and near-future forecasting~(i.e., after 1, 3, and 10 hours respectively), in both 2D and 3D scenarios. As a result, ImgSensingNet can achieve the best inference accuracy~(referred as the lowest RMSE in the table) in both real-time inference and future AQI forecasting. Even with high accuracy, the competitors may either lack the ability of future prediction~(e.g., ARMS) or 3D inference~(e.g., LSTM Nets).

\subsection{Energy Efficiency}
The energy-efficiency is analysed in two aspects: (1) the consumption of aerial UAV sensing, and (2) the consumption of ground WSN sensing. We choose AQNet~\cite{aqnet} that has similar components (using UAV and ground WSN) for comparison in the two aspects, respectively.

\textbf{Consumption of Aerial Sensing:} We set up experiments by comparing the normalized system consumption in monitoring tasks with different coverage spaces. As shown in Fig.~\ref{fig:consume compare uav}, ImgSensingNet uses UAV that does not suffer from both the \emph{load} and \emph{loitering} consumptions, hence can greatly save the battery. Compared to AQNet system with different loitering time $t$ for data sensing, ImgSensingNet consumes about 50\% less energy than that of AQNet, with different coverage space. Thus, the energy-efficiency of the proposed system is demonstrated.

\textbf{Consumption of Ground Sensing:} We further study the normalized consumption of ground sensing using the same method. We compare one day's consumption of all ground devices within different coverage spaces, using the same detection time and uploading time for each method. Fig.~\ref{fig:consume compare wsn} presents the experimental results. When $\text{\emph{JE}}=0$, our ground sensing achieves the maximum consumption, which still slightly outperforms AQNet system. As $\text{\emph{JE}}=0.5$, the normalized consumption of the WSN significantly reduces to only 53\%, which again validates the energy-efficiency of ImgSensingNet.

\subsection{Wake-up Mechanism}
We analyse the impact of $r$ on wake-up mechanism in two aspects: (1) the average number of devices that wake up each time, and (2) the average computing time for devices selection. We vary the number of devices as 30 and 100, and set $k=5$. For each instance, we perform 1000 independent runs to get the average values.

\textbf{Average Number of Wake-up Devices:} As shown in Fig.~\ref{fig:wakeup}(a)(c), we plot the average number of wake-up devices with different values of $r$, by setting $\text{\emph{JE}}=0$ as an invariant. The number of selected devices decreases monotonically when $r$ increases. Specifically, when we choose $r=300$ m, the average number of wake-up devices can greatly reduce to less than 50\% of total devices (e.g., 38.5 on average when there are 100 devices in total). Thus, by choosing a proper $r$, the number of wake-up devices greatly scales down, which is energy efficient.

\textbf{Average Runtime of Wake-up Mechanism:} Further, we study the runtime for obtaining the set of wake-up devices each time. As shown in Fig.~\ref{fig:wakeup}(b)(d), the runtime also decreases with a greater $r$. Specifically, the average running time is about 0.01 s when there are 30 devices in total. When there are more devices, the computation time will increase, but it is still completed in real-time~(about 1s in Fig.~\ref{fig:wakeup}(d)).

\begin{figure}[!t]
\small
\centering
\includegraphics[width=0.5\textwidth]{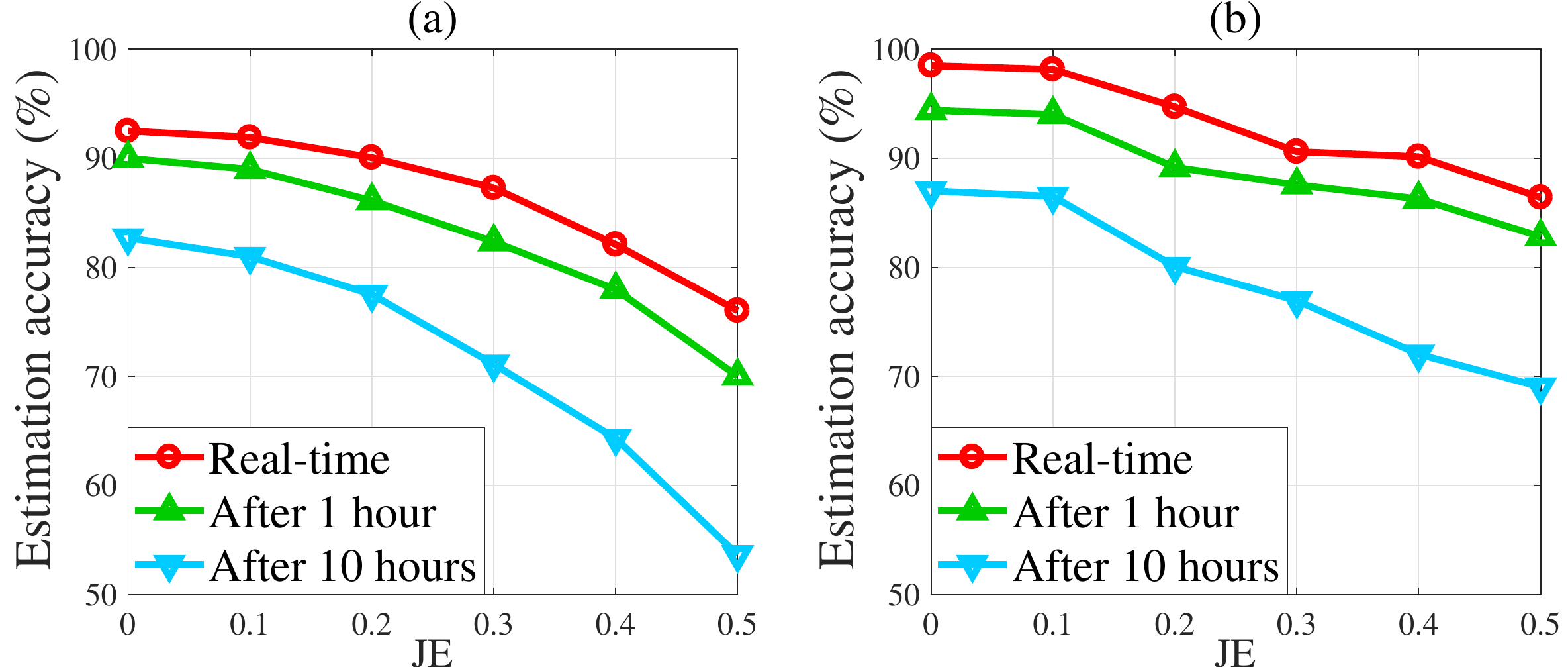}
\caption{The system estimation accuracy in real-time, after 1 hour, and after 10 hours, versus different \emph{JE}. (a) with 30 total devices; (b) with 100 total devices.}
\label{fig:diff je acc}
\vspace{-0.3cm}
\end{figure}

\begin{figure}[!t]
\small
\centering
\includegraphics[width=0.5\textwidth]{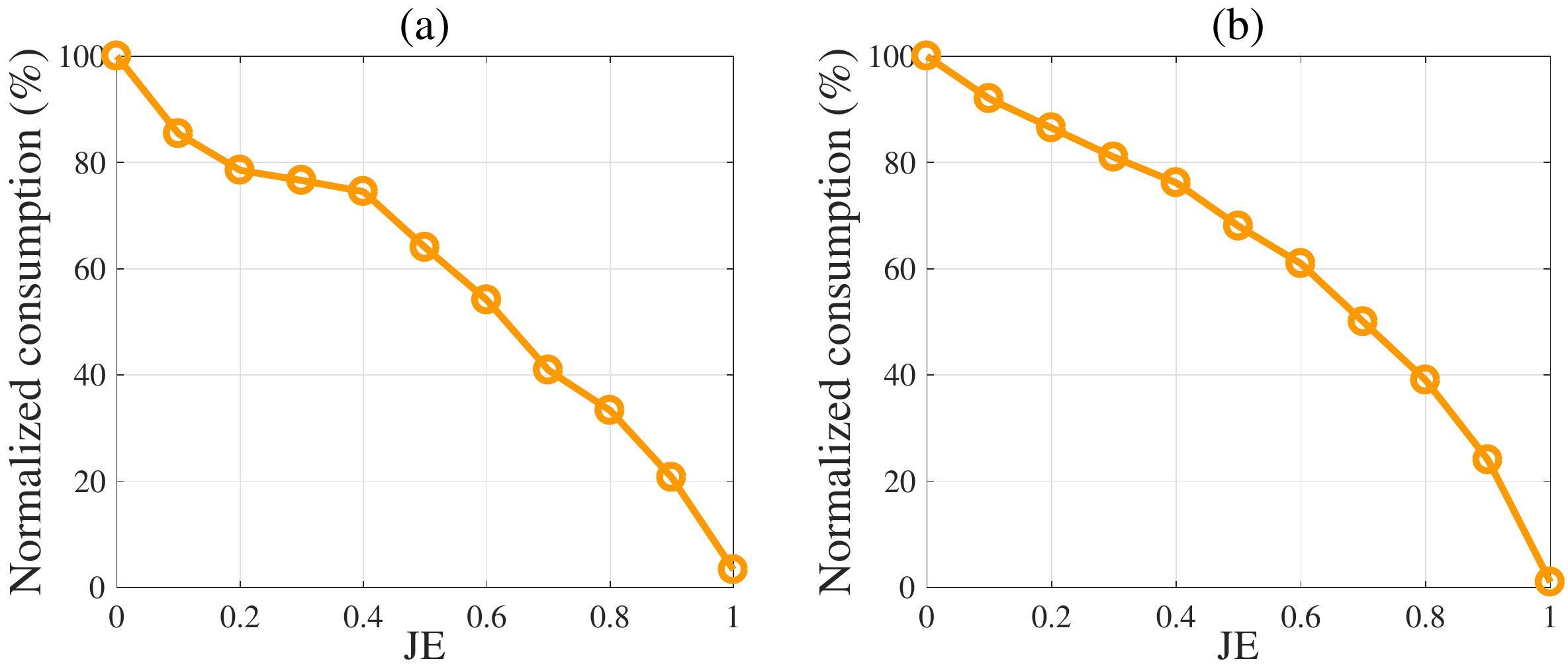}
\caption{The system normalized energy consumption versus different \emph{JE}. (a) with 30 total devices; (b) with 100 total devices.}
\label{fig:diff je consume}
\vspace{-0.3cm}
\end{figure}

\subsection{Impacts of Different Joint Estimation Errors}
In this section, we investigate the impacts of different \emph{JE} values on ImgSensingNet, in three aspects as (1) estimation accuracy, (2) energy consumption, and (3) working durations, respectively.

\textbf{Estimation Accuracy:} As shown in Fig.~\ref{fig:diff je acc}, the estimation accuracy gradually decreases when \emph{JE} increases. From the figure, we can see that ImgSensingNet achieves high accuracy in both real-time inference and future forecasting. Moreover, the system can achieve higher inference accuracy when there are more devices deployed.

\textbf{Normalized Energy Consumption:} In Fig.~\ref{fig:diff je consume} we report the relationship between energy consumption and different \emph{JE} values. By comparing Fig.~\ref{fig:diff je consume}(a) and Fig.~\ref{fig:diff je consume}(b), the energy consumption scales down as \emph{JE} increases, while a more stable procedure is obtained when there are more devices.

\begin{figure}[!t]
\small
\centering
\includegraphics[width=3in]{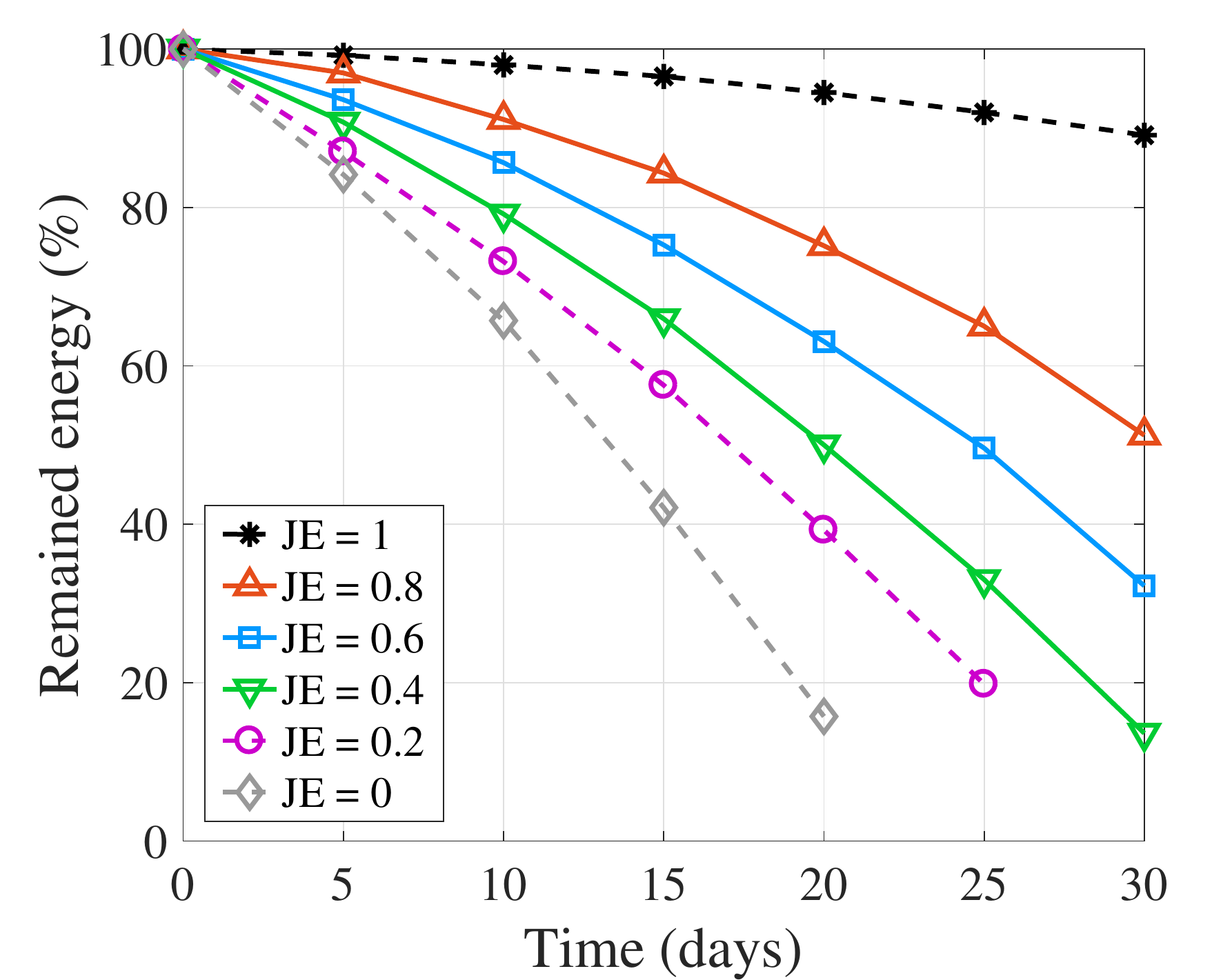}
\caption{The impact of different $\text{\emph{JE}}$ on system working durations.}
\label{fig:durations}
\vspace{-0.3cm}
\end{figure}

\textbf{System Working Durations:} In Fig.~\ref{fig:durations} we study the impacts of different $\text{\emph{JE}}$ on system working durations, over a fixed area inside Peking University. It is shown that ImgSensingNet can guarantee a long battery duration for more than one month when $\text{\emph{JE}}\ge0.4$, which greatly outperforms state-of-the-art systems. As \emph{JE} decreases, the monitoring overhead would increase, while it can also bring high inference accuracy. Hence, there exists a tradeoff between consumption and accuracy caused by different \emph{JE}, which needs to be further studied.

\subsection{Impact of Degree of Air Pollution}
\begin{figure}[!t]
\small
\centering
\includegraphics[width=0.5\textwidth]{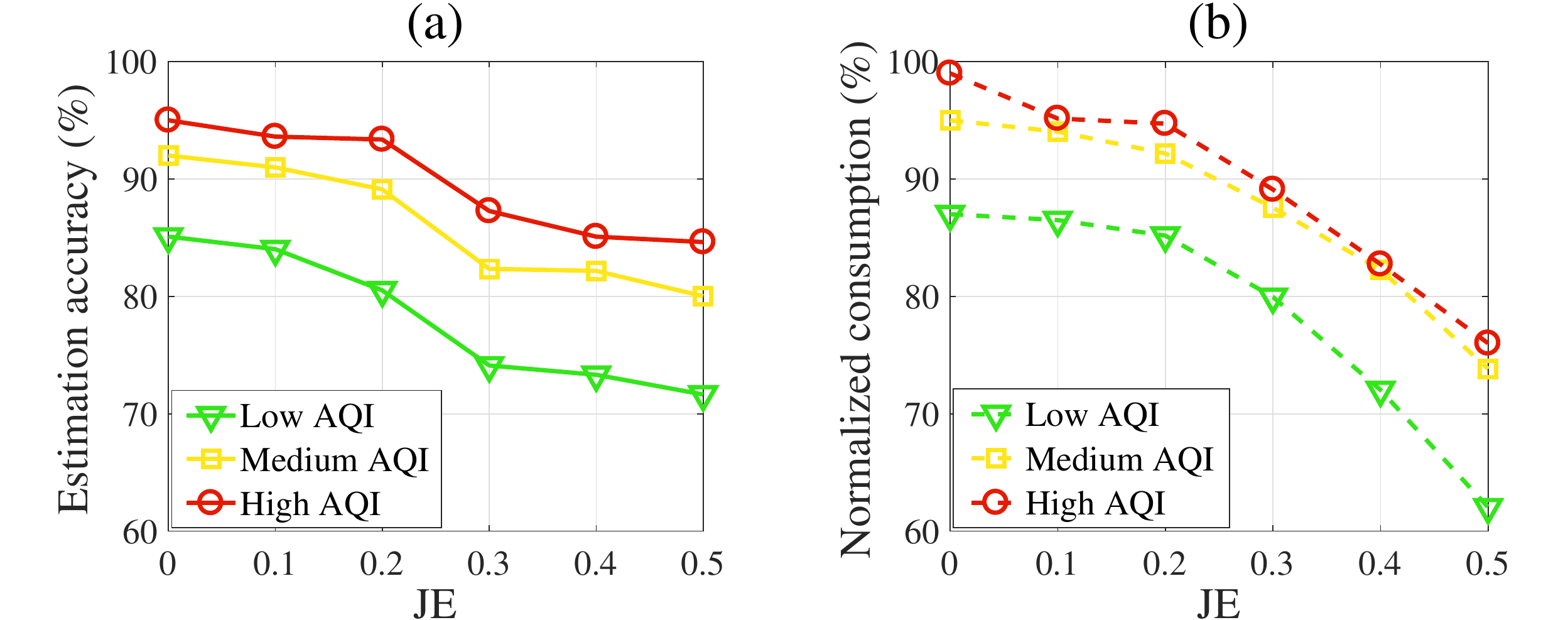}
\caption{The impact of degree of air pollution: (a) the estimation accuracy; (b) the normalized consumption.}
\label{fig:diff aqi}
\vspace{-0.3cm}
\end{figure}

In Fig.~\ref{fig:diff aqi}, we study the impact of the degree of air pollution on ImgSensingNet. We first manually divide our dataset into three degrees as slightly, moderately and highly polluted~(i.e., $\text{AQI}\le 50$, $50<\text{AQI}<200$ and $\text{AQI}\ge200$), and evaluate the performance of our model separately.

\textbf{Estimation Accuracy:} In Fig.~\ref{fig:diff aqi}(a) we compare the inference accuracy when AQI value varies. As a result, out system performs the best when $\text{AQI}\ge200$. Moreover, the performance tends to be better when AQI value is higher, as most devices are scheduled to sleep when air quality is good.

\textbf{Normalized Energy Consumption:} We further study the normalized energy consumption in different AQI degrees with various values of \emph{JE}. From Fig.~\ref{fig:diff aqi}(b), we can see that our system maintains the lowest consumption when AQI value is low, which again validates the energy-efficiency of the wake-up mechanism. By comparing Fig.~\ref{fig:diff aqi}(a) and (b), the tradeoff can also be illustrated.

\subsection{Tradeoff between Accuracy and Consumption}
\begin{figure}[!t]
\small
\centering
\includegraphics[width=3in]{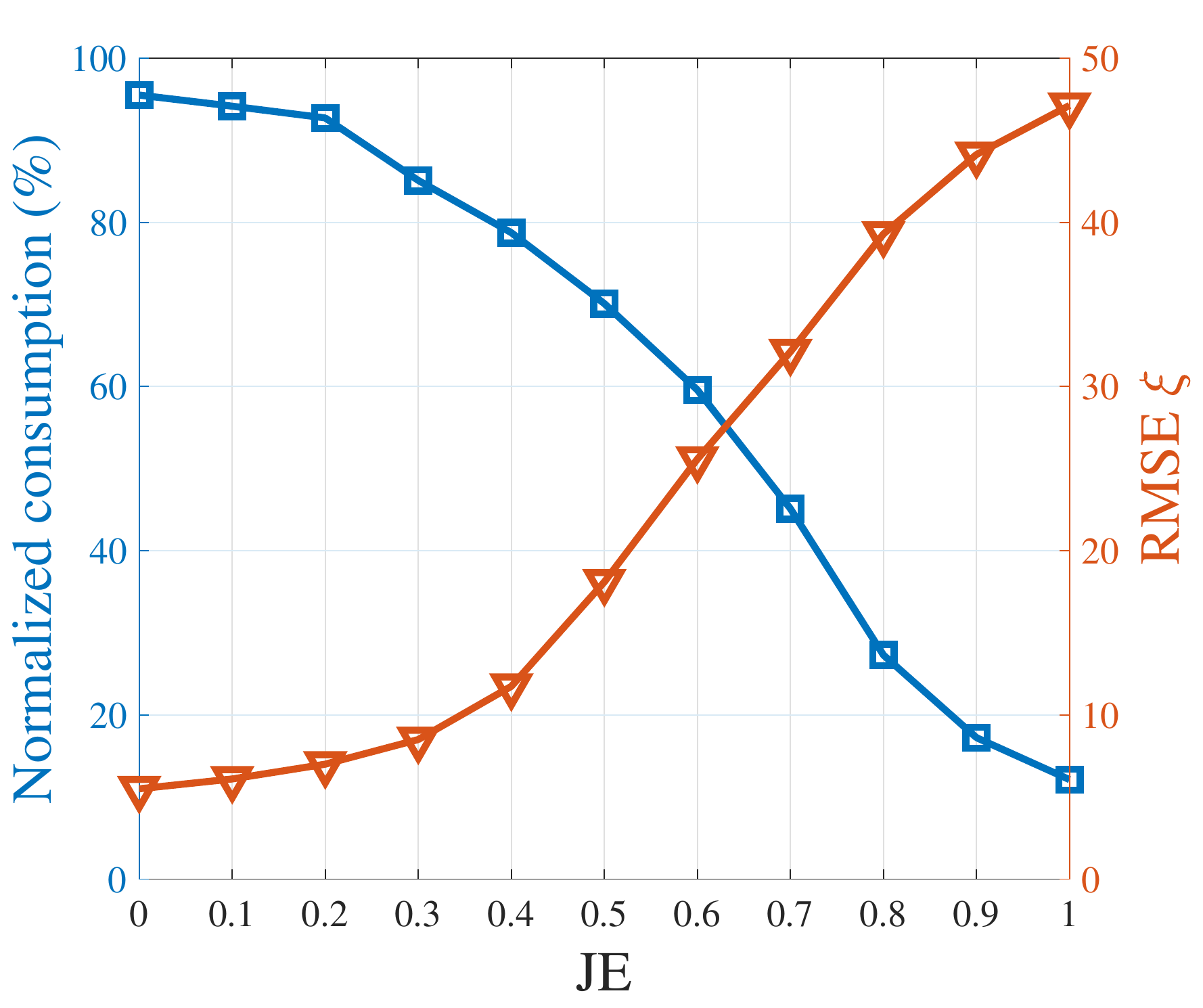}
\caption{The tradeoff between system power consumption and inference accuracy.}
\label{fig:tradeoff}
\vspace{-0.3cm}
\end{figure}

In Fig.~\ref{fig:tradeoff}, an inherent tradeoff between system consumption and inference accuracy is illustrated, versus \emph{JE}.
As \emph{JE} becomes higher, the average inference error grows rapidly while consumption can drop fairly. Given the average error, for example, when $\text{RMSE}$ is $25$, the corresponding $\text{\emph{JE}}=0.6$, which indicates the power consumption can be reduced to as little as $60\%$.
Hence, by choosing proper \emph{JE} value, the measuring cost can greatly scale down.

\section{Conclusion}
This paper presents the design, technologies and implementation of ImgSensingNet, a UAV vision guided aerial-ground AQI sensing system, to monitor and forecast the air quality in a fine-grained manner. We first utilize vision-based aerial UAV sensing for AQI scale inference, based on the proposed haze-relevant features and 3D CNN model. Ground WSN sensing are then used for accurate AQI inference in spatial-temporal perspectives using an entropy-based model. Further, an energy-efficient wake-up mechanism is designed to greatly reduce the energy consumption while achieving high inference accuracy. ImgSensingNet has been deployed on two university campuses for daily monitoring and forecasting. Experimental results show that ImgSensingNet outperforms state-of-the-art methods, by achieving higher inference accuracy while best reducing the energy consumption.

\end{document}